\documentclass{aastex631}

\definecolor{cobalt}{RGB}{33,70,139}
\definecolor{pharaoh}{RGB}{201,162,39}
\definecolor{nile}{RGB}{14,124,134}

\usepackage{enumerate}
\usepackage{mathtools}

\usepackage{amsmath}

\journalinfo{Accepted for publication in ApJ\\
  Received June 30, 2026; Revised October 7, 2026; Accepted October 14, 2026\\
  \doi{10.3847/1538-4357/aeb548}}

\renewcommand{\edit}[2]{#2}

\begin{document}

\title{A Generalized Framework for Rapid Polarization Response Generation in Wide Field-of-View Compton Polarimeters}

\author[0009-0003-3658-3316]{\edit1{Mohamed Gamal} Zaid}
\affiliation{University of New Hampshire, Space Science Center, Durham, 03824, New Hampshire, USA}

\author[0000-0003-0441-4959]{Merlin Kole}
\affiliation{University of New Hampshire, Space Science Center, Durham, 03824, New Hampshire, USA}

\author[0000-0001-8186-5978]{Mark McConnell}
\affiliation{University of New Hampshire, Space Science Center, Durham, 03824, New Hampshire, USA}
\affiliation{Southwest Research Institute, Dept. of Earth, Oceans, and Space, 03824, New Hampshire, USA}

\author[0000-0003-3331-3794]{Fabio Muleri}
\affiliation{INAF-IAPS, Via del Fosso del Cavaliere 100, 00133, Roma, Italia}

\author[0000-0002-5963-1494]{Hancheng Li}
\affiliation{Department of Astronomy, University of Geneva, 16 Chemin d'Ecogia, Versoix, CH-1290, Switzerland}

\begin{abstract}
\edit1{Polarimetry is the most direct probe of the unsettled radiation mechanism and magnetic field structure of gamma-ray burst (GRB) prompt emission at the energies where it peaks, tens of keV to about 1 MeV. Compton polarimeters use the azimuthal scattering angle distribution (ASAD), a 180 degree modulation on axis. GRB measurements have largely relied on the} on-axis \edit1{modulation factor $\mu_{100}$. For a wide field-of-view planar polarimeter, GRBs arrive off axis, where} the Klein-Nishina \edit1{projection adds harmonics to the ASAD. \citet{fabio} described this projection and evaluated it in a} low-energy \edit1{expansion for X-rays. We evaluate it numerically without approximation over the whole prompt-emission energy band for any incidence} and \edit1{polarization angles, quantify the off-axis harmonics a single modulation factor cannot capture, and verify the model} against GEANT4 simulations of an ideal polarimeter\edit1{. We also include the effect of the detection threshold on the ASAD. With POLAR as the case study, we present two modes that build the full polarization response with at least 95 percent fewer Monte Carlo simulations. Mode 1 takes only the unpolarized simulation per energy and recovers the polarization measured with POLAR's} full GEANT4 \edit1{responses for all but the brightest GRBs, where it underestimates the polarization degree. Mode 2 adds two polarized simulations per energy at distinct angles and reproduces the simulated ASADs to their statistical noise and the polarization of every GRB tested. Both modes are released as open-source software for any planar Compton polarimeter.}
\end{abstract}


\section{Introduction} \label{sec:intro}
The \edit1{radiation mechanism and the magnetic field structure of gamma-ray burst (GRB) prompt emission remain unsettled, and polarization is the most direct observable that separates the candidate models \citep{Gill2021}. That emission peaks at hard X-ray and gamma-ray energies, from tens of keV to about 1 MeV \citep{vonKienlin2020}, the band this work addresses throughout. The} last two decades have seen a rapid growth of polarimetry in high energy astrophysics. This was initiated by a range of polarization measurements of \edit1{GRBs} using instruments like RHESSI \citep{CoburnBoggs2003, RutledgeFox2004}, BATSE \citep{Willis2005} and those on INTEGRAL \citep{McGlynn2007, Gotz2009, Gotz2013, Gotz2014} and followed by several measurements of the Crab and Cygnus~X-1 at gamma-ray energies by INTEGRAL SPI and IBIS \citep{Dean2008, Forot2008}. This was followed by a wealth of measurements at X-ray energies thanks to instruments like PoGOLite, PoGO+, POLAR-Light, XL-Calibur and most recently IXPE \citep{Weisskopf2022}. In addition, instruments like GAP \citep{Yonetoku2011GAP, Yonetoku2011detection, Yonetoku2012}, POLAR and AstroSAT CZTI \citep{astrosat1, astrosat2} have, to date, produced tens of GRB polarization measurements, thereby paving the way for future large scale GRB polarimeters such as LEAP, Daksha \edit1{\citep{Bhalerao2024Daksha}}, COSI and finally POLAR-2 which will contain both a gamma-ray polarimeter (the High energy Polarization Detector (HPD)) and the first dedicated X-ray polarimeter for transients (the Low energy Polarization Detector (LPD)) \citep{Kole2025POLAR2}.

In parallel to this growth in measurements, progress has been made regarding the analysis in this field although standard, well tested analysis methods remain elusive. While the initial polarization measurements suffered from significant issues in the analysis, an overview of which is discussed in \cite{Gill2021}, a first step towards standardized analysis methods has been produced \citep{Burgess2019} and the first public polarization analysis tools have been published in recent years \citep[see, e.g.,][]{Kole2020}. Because the dominant photo-detector interaction is different at different energies, photo-absorption below a few tens of keV and Compton scattering above, the analysis approaches for soft $X-$ray and hard $X-$ray/$\gamma-$ray polarimetry have evolved largely independently. The analysis at $X-$ray energies is now well established and makes use of Stokes parameters. \edit1{Most} $X-$ray polarimeters are pointing instruments observing on-axis sources, for which the Stokes formalism is naturally suited; the off-axis complications central to this work do not arise there. While the use of Stokes parameters for polarimetry has long been used in optical and radio astronomy, at X-ray energies the discrete nature of the photons requires the calculation of Stokes parameters for individual photons. A method to deal with this in a mathematically sound manner was presented in \citet{Kislat2015} and has since been followed in the majority of X-ray analyses. 

At hard X-ray and gamma-ray energies a more wide range of analysis methods has been employed. Initial analysis typically relied on the production of so called modulation curves, which are defined in more detail in the next section. These modulation curves are fitted using single harmonic function with a period of $180^\circ$. Examples of this type of analysis can be found in the AstroSat CZTI \citep{astrosat1, Chand2019, Sharma2019} and GAP \citep{Yonetoku2011detection, Yonetoku2012} GRB analysis. While this method can work well for on-axis observations, as is the case for the analyses presented by the INTEGRAL teams \citep{Bosnjak2014, Gotz2009, Gotz2013}, for off-axis polarization observations this method is not correct. The off-axis nature, although appearing trivial at first sight, alters the polarization signal from a simple, single component harmonic, into a complex harmonic function as first discussed in detail in \cite{fabio}. The work in \cite{fabio} covers both the low energy X-ray range, where photo-absorption is used to measure the polarization of incoming photons, as well as the hard X-ray range where Compton scattering is used. For the Compton scattering case, on which we will focus here, the dependency of the azimuthal Compton scattering angle on the polarization of the incoming photons is presented in detail. In particular the \edit1{dependence} of the angular distribution on the incoming direction of the photon is \edit1{derived in general, and the resulting integrals are evaluated in a first-order expansion in the photon energy. That expansion holds} up to several \edit1{tens} of keV, \edit1{the range of X-ray polarimeters, and its approximations fail at the energies of GRB prompt emission}. As the field of polarimetry is growing especially in the gamma-ray energy range for wide field of view instruments, we revisit the off-axis case here and \edit1{evaluate the same description without approximation across the energy range defined above}.

In addition, the method is tested on existing data, and the effect of some of the simplifications made in previous analyses are also presented. This regards specifically the standard method where the polarization is retrieved through fitting a harmonic function. \edit1{In} recent years, more advanced methods, such as using forward folding in spectro-polarimetry, have been developed \citep{Burgess2019, Kole2020}. Such methods rely on comparing the measured modulation curves with simulated ones and retrieve the polarization and spectral parameters through minimization. As a result, they do not rely on fitting the harmonic functions, but rather on a comparison of the harmonic functions in the data with those present in the simulated detector response\edit1{, so it matters} that these complex harmonic functions are well represented in the \edit1{simulated responses, and we compare them in detail here. This work delivers the off-axis azimuthal scattering angle distribution evaluated numerically and without approximation across that band, for any incidence and polarization angles, verified against GEANT4 simulations of an ideal polarimeter. It adds the effect of the detection threshold on the measured distribution, which every real instrument imposes, with POLAR as the case study. On that description we build the full polarization response of an instrument in two modes that differ in their input: mode 1 takes only the unpolarized simulation at each energy, and mode 2 adds two fully polarized simulations per energy at distinct polarization angles to carry the polarization-dependent effects of the instrument. Either mode needs at least 95 percent fewer Monte Carlo simulations than a full response. The released software builds either response for any planar Compton polarimeter from that small set of its own simulations.}

The paper is organized as follows: Section \ref{sec:comptonIntro} reviews the Compton scattering formalism for polarized photons\edit1{,} sets the notation used throughout\edit1{, and states the off-axis problem}. Section \ref{sec:comptonGen} develops the generalized off-axis framework based on numerical integration of the transformed Klein-Nishina cross-section, which \edit1{holds across the band defined above, and describes the ideal simulations used to verify it; the low-energy expansion it supersedes is recorded in Appendix \ref{app:lowE}}. Section \ref{sec:realInstrument} extends the framework to realistic instruments \edit1{through the error-function threshold gate (derived in Appendix \ref{app:gate}) and the treatment of instrumental effects, and introduces the two response modes of this work}. Section \ref{sec:app} describes the current approaches to measuring polarization based on \edit1{the} azimuthal scattering angle distribution and full GEANT4 simulated responses of the instrument \citep{GEANT4_2003}\edit1{, and how each response mode is constructed: mode 1 from the unpolarized simulation alone, and mode 2 from the unpolarized simulation and fully polarized simulations at two polarization angles per energy}. In Section \ref{VerifySim}, the analytical predictions are compared against the GEANT4-generated responses of an ideal polarimeter and of the POLAR instrument \citep{Produit2018} for a range of incidence angles and spectra\edit1{, and the two modes are compared against the full polarized grids}. Section \ref{sec:results} then \edit1{applies both response modes to realistic polarization analyses: the} real POLAR observation of GRB 170114A \edit1{and a grid of synthetic} GRBs with known input polarization parameters, quantifying the regime in which the \edit1{mode 1 assumption holds and showing that mode 2 reproduces the full simulated response at the posterior level}. Finally, Section \ref{sec:future} summarizes the results and discusses future implementations, enabled by the reduction in computational cost. \edit1{The software implementing both modes is described in Appendix \ref{app:menhag}.}

\section{Polarimetry using Compton Scattering}\label{sec:comptonIntro}

\subsection{The On-Axis Ideal Case}\label{subsec:comptonOnAxisIdeal}
Photons \edit1{in the hard X-ray and gamma-ray band of Section \ref{sec:intro}, from} tens of keV \edit1{to about 1 MeV} can interact with matter through several processes. In the low-Z materials, typically used in Compton polarimeters, the two dominant processes are photoelectric absorption, in which the photon is fully absorbed and a bound electron is ejected, and Compton scattering, in which the photon scatters inelastically off an electron, transferring part of its energy. Photons can also undergo Rayleigh scattering, an elastic process in which the photon is redirected by the atomic cloud without any energy transfer. Since Rayleigh scattering deposits no energy at the interaction site, it cannot be detected in a Compton polarimeter and is therefore irrelevant to the measurement. In the energy range considered in this work, Compton scattering is the dominant interaction and provides the physical basis for polarization measurements \citep{knoll, attix1986}.

Compton scattering affects the energy and momentum of both the photon and the emitted electron during interaction. The Compton scattering equation can be derived by applying conservation of relativistic energy and momentum for the collision between a photon and a free electron. This yields the Compton kinematics equation: 
\begin{equation}\label{compton_eq}
    \frac{E'_\gamma}{E_\gamma} = \frac{1}{1+\varepsilon (1-\cos{\theta})} \;\;\;\;\;\;\;\;
    \varepsilon=\tfrac{E_\gamma}{\text{511 keV}}
\end{equation}

Here $E_\gamma$ is the energy of the incoming photon, $E'_\gamma$ is the scattered photon energy, and $\theta$ is the Compton scattering angle, defined as the angle between the incident momentum $\vec{k}$ and the scattered momentum $\vec{k}'$. It is labeled $\theta_C$ in the following figures; the two symbols denote the same angle. Because the incident photon travels along the $-z_\gamma$ axis, $\theta$ is measured from $-z_\gamma$ by convention. For readers using standard spherical coordinates, it relates to the polar angle $\theta_s$ (measured from $+z_\gamma$) by $\theta=\pi-\theta_s$, equivalently $\cos\theta=-\cos\theta_s$. This convention is used consistently in all text, equations, and figures.

A full three dimensional description of the scattering process is given by the Klein-Nishina cross-section \citep{KleinNishina1929}. It provides a fully relativistic treatment of the scattering process, derived from the Dirac equation and Quantum Electrodynamics (QED), which naturally incorporates the polarization state of the incident photon. Consequently, the polarization state of the incident photon directly affects the kinematics of the scattering cross-section. Because the incident photon's oscillating electric field preferentially accelerates the target electron along the axis of polarization, the resulting scattering probability loses its azimuthal symmetry about the incident photon direction.

\begin{figure}[ht]
    \centering
    \includegraphics[width=0.4\linewidth]{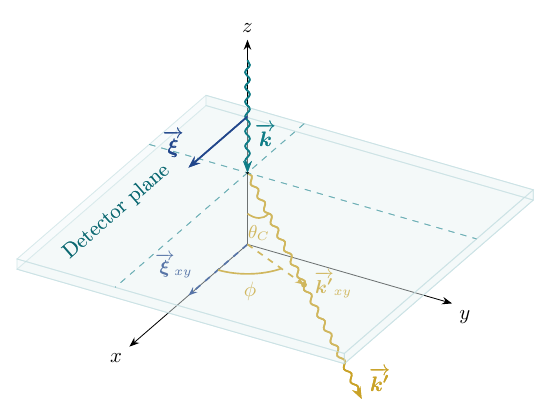}
    \caption{The idealized photon reference frame for Compton scattering. The incident photon momentum $\vec{k}$ is aligned with the negative $z-$axis, and the polarization vector $\vec{\xi}$ is along the positive $x-$axis. In this configuration, the Klein-Nishina azimuthal angle $\phi$ coincides with the standard spherical azimuthal coordinate.}
    \label{fig:photonFrame}
\end{figure}

The Klein Nishina differential cross section is given by: 
\begin{equation}\label{KN_eq}%
  {\frac {d\sigma }{d\Omega }}={\frac {1}{2}}r_{e}^{2}\left({\frac {E_\gamma'}{E_\gamma}}\right)^{2}\left[{\frac {E_\gamma}{E_\gamma'}}+{\frac {E_\gamma'}{E_\gamma}}-2\sin ^{2}(\theta )\cos ^{2}(\phi)\right] 
\end{equation}
Here $r_e$ is the classical electron radius\edit1{, ($r_e$ is given by $r_e=\frac{\alpha\hbar}{m_e c}\approx2.82\times10^{-15}$ m. Here $\alpha$ is the fine structure constant, $\hbar$ is the reduced Planck's constant, $m_e$ is the electron's rest mass and $c$ is the speed of light.) and} $\theta$ is the same as in equation \ref{compton_eq}. The angle $\phi$ is the azimuthal scattering angle defined as the angle between incident photon polarization vector $\vec\xi$ and the scattered photon direction in a plane perpendicular to incident photon direction as illustrated in Figure \ref{fig:photonFrame}. 

In a physical polarimetry experiment, measurements are recorded relative to a macroscopic instrument frame, typically a large area planar detector defined by an $xy-$plane. Figure \ref{fig:photonFrame} shows the simplified case where the intrinsic "photon frame" is perfectly aligned with this physical detector frame. Here, the photon's incident momentum direction $\vec{k}$ is aligned with the detector's negative $z-$axis, and its polarization vector $\vec{\xi}$ is aligned perfectly with the detector's positive $x-$axis. The scattering angle $\phi$ in this configuration identically matches the azimuthal angle of the vector $\vec{k'}$ in the standard spherical coordinate system. However, in practice, a physical detector will measure a more generalized case.

The Klein-Nishina cross-section dictates that polarized photons preferentially scatter in directions perpendicular to their incident polarization vector. For the specific configuration shown in Figure \ref{fig:photonFrame}, this corresponds to the angles $\phi=\pi/2\;,3\pi/2$, where Equation \ref{KN_eq} reaches its maximum. This implies that the azimuthal scattering distribution of the photons contains the necessary information to determine the bulk polarization of the incoming beam. By contrast, an incident unpolarized beam lacks a defined polarization vector and will scatter with equal probability in all azimuthal directions.

\begin{figure}[ht]
    \centering
    \includegraphics[width=0.5\linewidth]{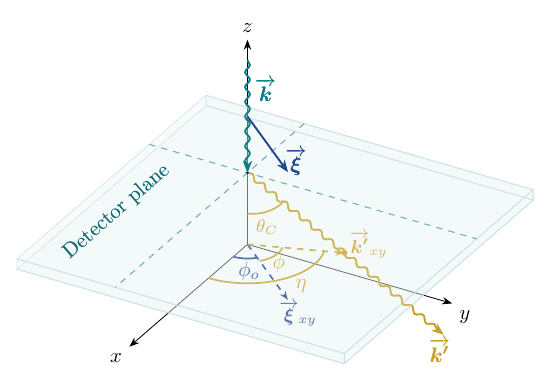}
    \caption{A generalized on-axis measurement in the detector frame. The incident photon momentum $\vec{k}$ remains along the negative $z-$axis, but the polarization vector $\vec{\xi}$ makes an angle $\phi_o$ with the detector's positive $x-$axis. The measured spherical azimuthal angle $\eta$ is related to the intrinsic Klein-Nishina scattering angle $\phi$ through a rigid planar rotation, $\eta=\phi+\phi_o$.}
    \label{fig:detectorFrame}
\end{figure}

\edit1{In a measurement} with a large-area planar detector \edit1{the} polarization vector $\vec{\xi}$ \edit1{lies at some angle in} the detector's $xy-$plane \edit1{(}$\vec{\xi}$ \edit1{must be orthogonal to $\vec{k}$, as required by Maxwell's equations); that angle from} the detector's positive $x-$axis is \textit{the polarization angle} $\phi_o$ \edit1{(Figure \ref{fig:detectorFrame}). This is a \emph{local} polarization angle defined in the detector frame. It should not be confused with the polarization angle defined in the source incidence frame, nor with the IAU North-to-East convention used to report astrophysical polarization angles; the relation between them is fixed by the source geometry introduced in Section~\ref{sec:comptonGen}. The Klein-Nishina angle} $\phi$ \edit1{is then} measured in the $xy-$plane \edit1{between the projections of $\vec{k}'$} and $\vec{\xi}$\edit1{, and} the azimuthal scattering angle \edit1{the detector records,} $\eta$, measured from the positive $x-$axis\edit1{, is the rigid rotation} $\eta=\phi+\phi_o$ \edit1{(to ensure mathematical consistency across all reference frames, all azimuthal angles, $\phi$, $\phi_o$ and $\eta$, are strictly defined as directed angles measured counter-clockwise in the $xy-$plane according to the right-hand rule); strict on-axis incidence is common in the laboratory and rare in orbit, and the off-axis case is treated in the following sections}.

\edit1{A polarimeter records $\eta$ for each scattered photon of a} beam incident along the negative $z-$axis. The resulting \textit{Azimuthal Scattering Angle Distribution (ASAD)} $\mathcal{N}(\eta)$ \edit1{follows} a $\cos^2{(\eta-\phi_o)}$ modulation, a Fourier second harmonic phase-shifted by $\phi_o$\edit1{, since the Klein-Nishina cross-section depends on $\cos^2{\phi}$; its first minimum determines the polarization angle. The} polar angle of $\vec{k'}$ here is $\pi-\theta$ where $\theta$ is the Compton scattering angle.

The expected shape of the ASAD resulting from a polarized beam incident along the negative $z-$axis of a detector can be seen through simple manipulations of equation \ref{KN_eq}. By considering only the dependence on the azimuthal scattering angle $\phi=\eta-\phi_o$, we can write 
\begin{equation}\label{onAxisDeriv}
\begin{split}
\frac{d\sigma}{d\phi}(\eta,\;\phi\edit1{_o}) &= \tilde{B} - \tilde{A}\cos ^{2}(\eta -\phi\edit1{_o}) = \tilde{B} - \tilde{A} \bigg[\tfrac{1}{2} \big(1+\cos(2(\eta-\phi_o))\big)\bigg]
= \underbrace{\left(\tilde{B} - \tfrac{\tilde{A}}{2}\right)}_{\text{Offset } B} - \underbrace{\tfrac{\tilde{A}}{2}}_{\text{Amplitude } A}\cos(2(\eta-\phi\edit1{_o})) \\&= B - A\cos(2(\eta-\phi_o))
\end{split}
\end{equation}

\begin{figure}[ht]
\centering
\includegraphics[width=0.6\textwidth]{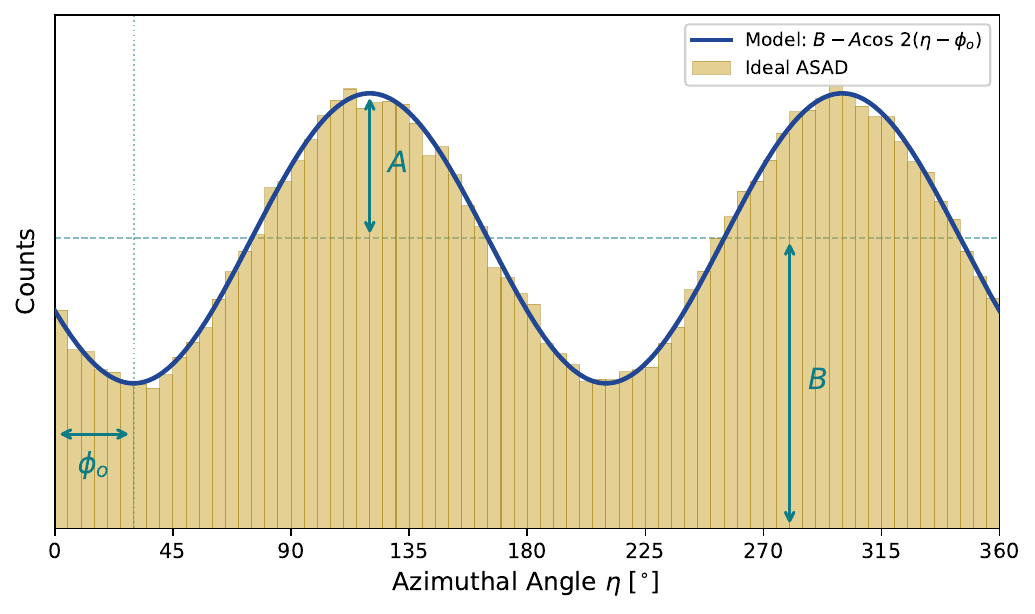}
\caption{A schematic illustration of the azimuthal scattering angle distribution (ASAD) for a polarized beam in an ideal detector, parameterized by the model of Equation~\ref{onAxisDeriv}: modulation amplitude $A$, offset $B$, and polarization angle $\phi_o$.}
\label{fig:ASAD_kn}
\end{figure}

The free parameters of this last equation $(A,\;B,\;\phi_o)$ quantify the polarization signature. The parameter B represents the average count rate (offset), and A represents the amplitude of the modulation. A typical "ideal" shape of the ASAD in terms of these parameters is shown in Figure \ref{fig:ASAD_kn}. 

The measured ASAD, $\mathcal{N}(\eta)$, is the distribution of recorded scattered photons as a function of the azimuthal scattering angle $\eta$. To quantify the strength of the polarization signal, we define the modulation factor $\mu$ from the maximum and minimum of this distribution, $\mathcal{N}_{\max}$ and $\mathcal{N}_{\min}$:
\begin{equation}\label{Eq:mod_eq1}
\begin{split}
\mu=\frac{N_{max}-N_{min}}{N_{max}+N_{min}}
\end{split}
\end{equation}
We can find $N_{max}$ from Equation \ref{onAxisDeriv} as $N_{max}=B+A$ and similarly $N_{min}=B-A$. This definition can then be simplified to $\mu=A/B$, which is simply the ratio of the amplitude of the variation to the average value of the ASAD. For a fully polarized beam measured by a perfect detector, this modulation reaches a theoretical maximum known as \edit1{$\mu_{\max}(E, \theta)$. (It can be referred to as $\mu_{100}$, but the literature reserves $\mu_{100}$ to refer to the measured modulation of a fully polarized beam simulation as a figure-of-merit for the instrument performance, so we label the theoretical limit of modulation as $\mu_{\max}(E,\theta)$.) It} is a fundamental limit that can be derived directly by comparing Equation \ref{KN_eq} (after expanding the $\cos^2{\phi}$ term and regrouping) to Equation \ref{onAxisDeriv} to yield the amplitude $A$ and the offset $B$, giving:  
\begin{equation}\label{Eq:mod_eq2}
\begin{split}
\mu\edit1{_{\max}}(E, \theta) = \frac{\sin^2\theta}{\frac{E'}{E} + \frac{E}{E'} - \sin^2\theta}
\end{split}
\end{equation}
Real instruments cannot measure this theoretical limit due to instrumental effects which can be captured in simulations as a part of the response in $(E,\theta)$ of that instrument. We define $\text{Response$(E,\theta)$}$ to denote the instrument's detection efficiency for a Compton event of incident energy $E$ and scattering angle $\theta$, i.e. the number of recorded coincidences per incident photon. We can then define the maximum measurable modulation factor for an instrument as the weighted average of the theoretical \edit1{$\mu_{\max}(E,\theta)$} integrated over the response as: 
\begin{equation}\label{Eq:mod_eq2_real}
\begin{split}
\mu_{100} = \edit1{\frac{\iint \mu_{\max}(E,\theta) \;\text{Response$(E,\theta)$}\; dE\; d\theta}{\iint \text{Response$(E,\theta)$}\; dE\; d\theta}}
\end{split}
\end{equation}
The measured polarization level is then calculated as $\Pi=\mu/\mu_{100}$. The formalism above applies only to on-axis sources, and real wide-FOV instruments observe sources at arbitrary incidence angles.

\subsection{The Off-Axis Reality (The Problem)}
It is evident from equation \ref{KN_eq} and \ref{onAxisDeriv} that the ASAD (after correcting for instrumental effects) is expected to have only a $180^{\circ}$ component. This has been the assumption in many, but not all, \edit1{GRB polarization measurements to date. Those analyses characterized the signal by a single $180^{\circ}$ modulation and its on-axis modulation factor $\mu_{100}$ regardless of the} source position. Examples of this can be found in \citep{astrosat1, astrosat2}. This assumption is, however, false when the source location is not along the viewing axis of the polarimeter. An X-ray polarimetry study proved that ASADs from such off-axis sources should include more components in the signal \citep{fabio}. This was also evident from ground calibrations of the POLAR instrument, where additional $360^{\circ}$ and $90^{\circ}$ components in the ASAD for off-axis source were measured by \cite{Kole2017POLARCalibration}. 

\begin{figure}[ht]
\centering
\includegraphics[width=0.9\textwidth]{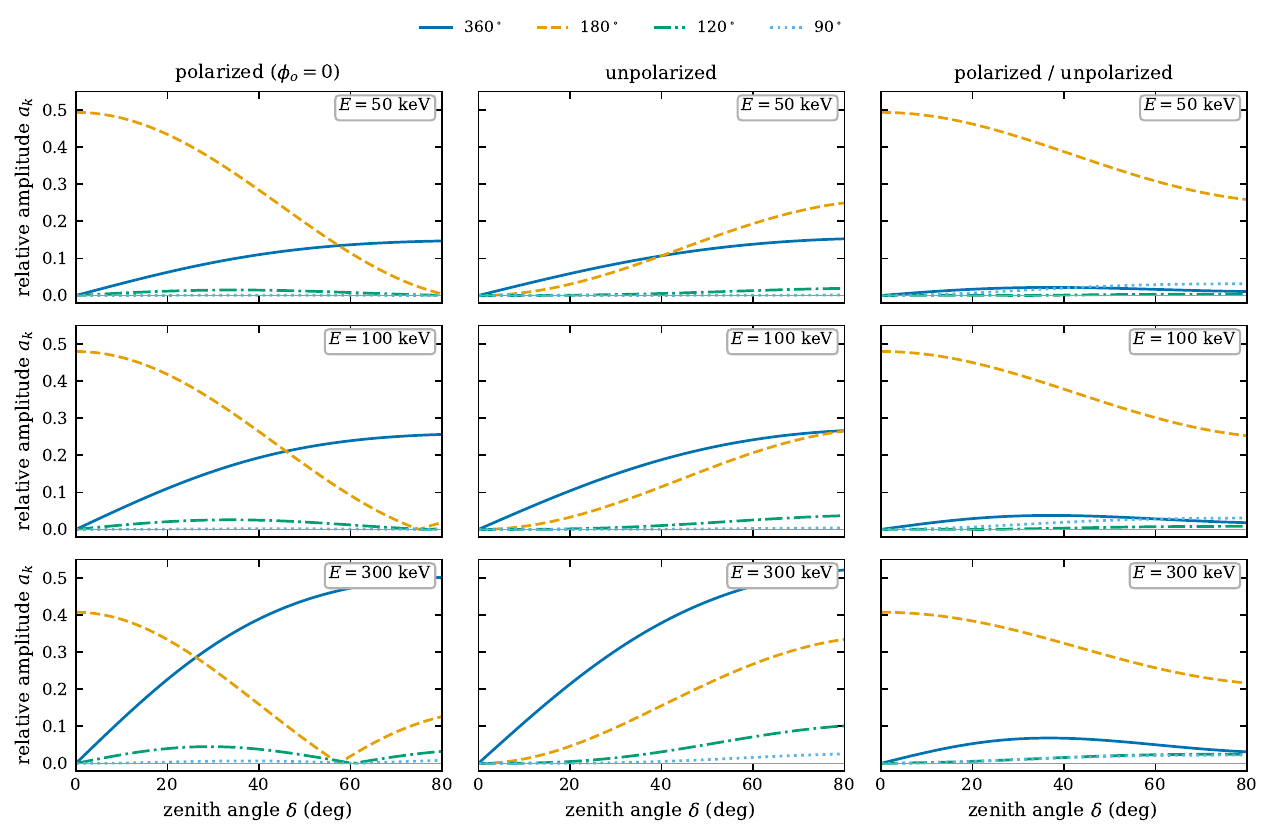}
\caption{\edit1{Relative} amplitudes \edit1{$a_k$} of the \edit1{first four} Fourier \edit1{components (periods $360^\circ$, $180^\circ$, $120^\circ$ and $90^\circ$)} of the ASAD \edit1{against the source zenith angle $\delta$} at \edit1{50, 100} and \edit1{300 keV, for the fully} polarized \edit1{ASAD at} $\phi_o=0$ (\edit1{left})\edit1{, the} unpolarized ASAD \edit1{(middle)} and \edit1{their ratio} (\edit1{right})\edit1{, computed with the framework of Section \ref{sec:comptonGen} with no threshold gate}.}
\label{fig:harmonicsAmplitudes}
\end{figure}

Expanding the ASADs in terms of Fourier harmonics is instructive in showing what components exist in the signal. To illustrate its importance we generated ASADs for different zenith positions $\delta\in[0^\circ,80^\circ]$ for energies $50, 100, \text{ and } 300 \text{keV}$ using our model which will be presented in the next section. And we show in Figure \ref{fig:harmonicsAmplitudes}, the relative amplitudes $a_k$ for the each of the first four harmonics for each ASAD. The relative amplitudes $a_k$ are measured with respect to the constant term in the Fourier expansion. It can be seen that for the polarized ASADs, the $180^\circ$ component is maximum only when we have on-axis incidence $\delta=0$. And the unpolarized ASADs have harmonics and are not just a constant value once photons are incident from an off-axis angle. In the typical context of classical polarization analysis, as presented in Section \ref{sec:classical}, the polarized ASAD is divided by an unpolarized ASAD which is argued to remove the geometric effects in the signal, but as shown in the right column of Figure \ref{fig:harmonicsAmplitudes},  it \edit1{does not} completely remove the geometric effects due to the projection of Klein-Nishina cross-section which are depicted as extra harmonics. Those additional Fourier components contaminate the $\cos^2{\phi}$ modulation that \edit1{traditional analysis fits, and Figure \ref{fig:harmonicsAmplitudes} quantifies the part of the off-axis signal that a single modulation factor cannot capture. Assuming a single} $180^\circ$ modulation component for off-axis sources \edit1{therefore} biases the recovered polarization, \edit1{which prompted} this study.

\section{Generalized Description of Compton Scattering}\label{sec:comptonGen}
To quantify the additional harmonics identified in the previous section, we derive a generalized form of the Klein-Nishina equation valid for an arbitrary source direction and polarization angle. The goal is to obtain a description of the azimuthal scattering angle distribution, in the detector frame and as a function of $\eta$, that accounts for the two rotations defining the source location and the rotation defining the polarization angle. The angles $\theta$ and $\phi$ in the Klein-Nishina equation are defined with respect to the photon frame which we denote by $x_\gamma y_\gamma z_\gamma$. When the coordinates of the incident photon frame align with those of the detector ($x_d y_d z_d$), only the typical second-order harmonic is observed in the ASAD. However, for a general incident direction the two frames would not align and we expect additional harmonics. This is because the photon polarization vector, by definition, lies in the plane perpendicular to the photon momentum vector. Consequently, if a photon is incident off-axis, projection effects will distort the typical $180^{\circ}$ component. 
\subsection{Coordinate Transformation}\label{subsec:coordTrans}

Here, we will utilize the spherical coordinate representation in both the photon frame $x_\gamma y_\gamma z_\gamma$ and the detector frame $x y z$. The description relies on describing the direction of the scattered photon momentum vector $\vec{k'}$ in terms of spherical coordinates. In the photon frame, $\theta$ and $\phi$ follow the convention established in Section \ref{subsec:comptonOnAxisIdeal} (Figure \ref{fig:photonFrame}). In the detector frame, the corresponding transformed polar and azimuthal angles are $\vartheta$ and $\varphi$, and the measured azimuthal angle is $\eta$. All angles used in the following derivation are summarized in \edit1{Appendix \ref{app:definitions}}. An example of a simple transformation from the photon frame to the detector is when the photon is rotated around the $z-$axis with angle $\phi_o$ (the polarization angle). This rotation $R_{\phi_o}^z$ makes the angle $\eta$ related to $\phi$ through \edit1{$\eta=\phi_o+\phi$} as shown in figure \ref{fig:detectorFrame}. For an arbitrary source location i.e., not on-axis, the transformation from the photon frame to the detector frame requires two rotations as follows: for zenith angle $\delta$ through $R_\delta^y$ and for azimuth angle $\lambda$ through $R_\lambda^z$.

\begin{figure*}[ht!]
    \gridline{
        \fig{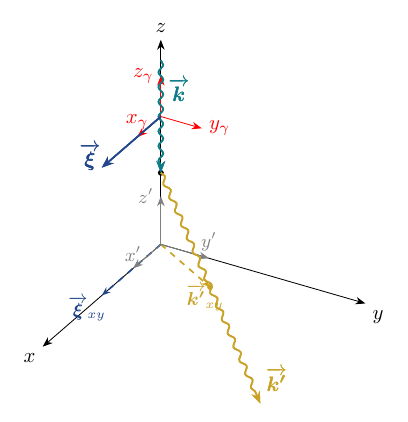}{0.32\textwidth}{(a) Initial orientation of the photon frame (denoted by $x_\gamma y_\gamma z_\gamma$) when aligned with the detector frame (denoted by $xyz$) as described in Figure \ref{fig:photonFrame}.}
        \fig{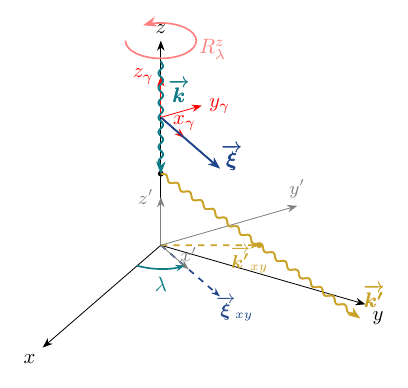}{0.32\textwidth}{(b) Step 1: Rotation by azimuth $\lambda$ around the $z-$axis, the photon frame is now aligned with $x'y'z'-$axes}
        \fig{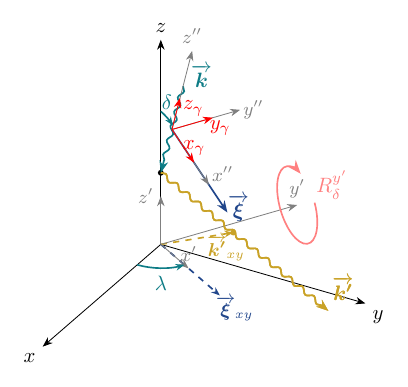}{0.32\textwidth}{(c) Step 2: Rotation by zenith $\delta$ around the $y'-$axis, the photon frame is now aligned with $x''y''z''-$axes}
    }
    \gridline{
        \fig{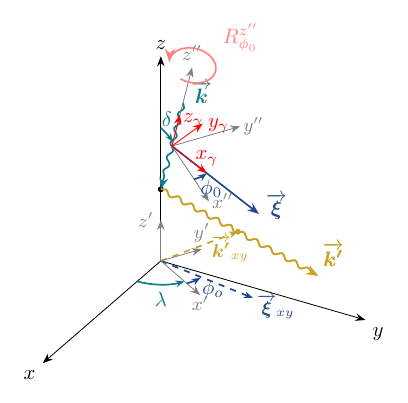}{0.4\textwidth}{(c) Step 3: Rotation by polarization angle $\phi_o$ around the $z''-$axis. This is the final orientation displaying all transformation parameters ($\lambda, \delta, \phi_o$).}
        \fig{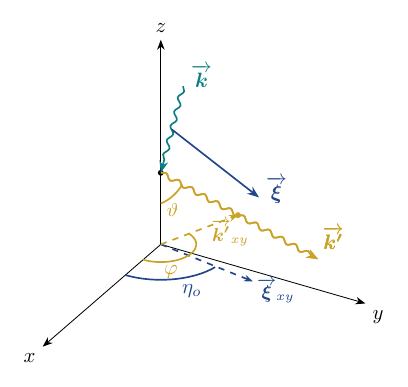}{0.4\textwidth}{(e) The final orientation with decluttered geometry highlighting the scattered photon's spherical coordinates ($\vartheta, \varphi$).}
    }
    \caption{The sequential coordinate transformations required to align the detector frame with the photon reference frame for off-axis incidence. Panel (a) shows the photon reference frame aligned with the detector frame. Panel (b) and (c) illustrate the subsequent rotations by azimuth angle $\lambda$ zenith angle $\delta$ and polarization angle $\phi_o$, respectively. Panel (d) shows the final orientation with all three transformation parameters. Panel (e) shows the spherical coordinates for the scattered photon vector $\vec{k'}$ in terms of $\vartheta,\;\varphi$ which are used in the generalized description of the ASAD in Equation \ref{eq:integralsTransformed}.}
    \label{fig:coordinate_transformations}
\end{figure*}

We first write the differential Compton scattering distribution in the photon frame, $\mathcal{D}^{\mathrm{pol}}_\gamma(\theta,\phi,\varepsilon)$ and $\mathcal{D}^{\mathrm{unp}}_\gamma(\theta,\varepsilon)$ the Klein-Nishina cross-section per unit solid angle (Equation~\ref{KN_eq}). Here the superscript denotes the polarized or unpolarized beam and the subscript $\gamma$ denotes the photon frame. Integrating over the polar angle and normalizing yields the azimuthal distribution $\Phi^{\mathrm{pol/unp}}_\gamma(\phi,\varepsilon)$. Finally, applying the frame rotations re-expresses this in the detector's azimuthal angle, $\Phi(\varphi)\equiv\Phi(\eta)$, which is the quantity a polarimeter measures. We start with equation \ref{KN_eq} and use equation \ref{compton_eq} to substitute for $E'_{\gamma}/E_{\gamma}$ as follows: 
\begin{equation}\label{eq:expandedKN_Pol}
    {\frac {d\sigma }{d\Omega }}\bigg|^{\text{pol}}_{\gamma}=\mathcal{D}^{\text{pol}}_{\gamma}(\theta,\phi, \varepsilon) = \tfrac{1}{2}r_e^2 \Bigg[\frac{1}{1+\varepsilon(1-\cos{\theta})} + \frac{1}{[1+\varepsilon(1-\cos{\theta})]^3} - \frac{2\sin^2{\theta}\cos^2{\phi}}{[1+\varepsilon(1-\cos{\theta})]^2}\Bigg]
\end{equation}
The "pol" superscript refers to the polarized portion of the beam. And for the unpolarized portion of the beam, $\cos^2{\phi}$ is replaced by its azimuthal average, $\langle \cos^2{\phi} \rangle=\tfrac{1}{2}$, yielding: 
\begin{equation}\label{eq:expandedKN_Unp}
    {\frac {d\sigma }{d\Omega }}\bigg|^{\text{unp}}_{\gamma}=\mathcal{D}^{\text{unp}}_{\gamma}(\theta, \varepsilon) = \tfrac{1}{2}r_e^2 \Bigg[\frac{1}{1+\varepsilon(1-\cos{\theta})} + \frac{1}{[1+\varepsilon(1-\cos{\theta})]^3} - \frac{\sin^2{\theta}}{[1+\varepsilon(1-\cos{\theta})]^2}\Bigg]
\end{equation}
Since $\mathcal{D}$ describes the distribution over a solid angle $d\Omega$, we integrate over the full range of $\theta$ to obtain the azimuthal angular distributions and to normalize this distribution, it is divided by the integration over the entire solid angle $\Omega$, i.e.
\begin{equation}\label{eq:integrals}
    \Phi^{\text{pol}}_{\gamma}(\phi, \varepsilon) = \frac{\int_0^{\pi}\mathcal{D}^{\text{pol}}_{\gamma}(\theta,\phi, \varepsilon)\sin{\theta} d\theta}{\int_0^{2\pi}\int_0^{\pi}{\mathcal{D}^{\text{pol}}_{\gamma}(\theta,\phi, \varepsilon)\sin{\theta} d\theta d\phi}}
    \;\;\;\;\;\;\;\;
    \Phi^{\text{unp}}_{\gamma}(\varepsilon) = \frac{\int_0^{\pi}\mathcal{D}^{\text{unp}}_{\gamma}(\theta, \varepsilon)\sin{\theta} d\theta}{\int_0^{2\pi}\int_0^{\pi}{\mathcal{D}^{\text{unp}}_{\gamma}(\theta, \varepsilon)\sin{\theta} d\theta d\phi}}
\end{equation}
This is now a representation of the normalized distribution $\Phi(\phi)\equiv\Phi(\eta)$ i.e., the modulation curve as viewed in the photon frame when it is anti-aligned with the detector frame. Equation \ref{eq:integrals} on the right shows that the expected unpolarized distribution for an on axis source and would be a constant for all values of azimuthal scattering angle as outlined earlier. However, for a general photon direction we need to transform the photon frame to anti-align with the detector frame. The transformation would consist of a series of rotations that correspond to the source photon's degrees-of-freedom, i.e., the azimuthal angle $\lambda$, the zenith angle $\delta$ and the polarization angle $\phi_o$ (Figure \ref{fig:coordinate_transformations}). We now express the direction of the scattered photon in both frames using spherical coordinates. Following the convention established in Section \ref{subsec:comptonOnAxisIdeal} ($\theta$ measured from $-z_\gamma$), the photon-frame $z$-component of the scattered photon is $z_\gamma=-\cos\theta$, as used in Equation \ref{eq:photonFrameAxes}. Throughout, a subscript $\gamma$ denotes a quantity expressed in the incident-photon frame. The incident photon frame unit basis vectors are given by: 
\begin{equation}\label{eq:photonFrameAxes}
    x_\gamma=\sin{\theta}\cos{\phi},\;\;\;y_\gamma=\sin{\theta}\sin{\phi},\;\;\;z_\gamma=-\cos{\theta}
\end{equation}
Similarly, the detector frame basis unit vectors are labeled by: 
\begin{equation}\label{eq:detectorFrameAxes}
    x=\sin{\vartheta}\cos{\varphi},\;\;\;y=\sin{\vartheta}\sin{\varphi},\;\;\;z=-\cos{\vartheta}
\end{equation}
The rotations are now applied as follows: 
\begin{equation}\label{eq:rotatingToPhotonFrame}
\begin{pmatrix}
    x_\gamma\\y_\gamma\\z_\gamma
\end{pmatrix}
= \mathcal{R}^{z''}_{\phi_o} \mathcal{R}^{y'}_\delta \mathcal{R}^{z}_\lambda
\begin{pmatrix}
    x\\y\\z
\end{pmatrix}\\
\end{equation}
After the first rotation by $\lambda$ around the $z-$axis, the coordinate system is now $x' y' z'$. Then the second rotation by $\delta$ around the $y'-$axis, the coordinate system is now $x'' y'' z''$. And the final rotation by $\phi_o$ around the $z''-$axis, the coordinate system now matches $x_\gamma y_\gamma z_\gamma$.

Accordingly, the rotation matrices are defined as follows: 
\begin{equation}\label{eq:rotationMatrices}
    \mathcal{R}^{z}_\lambda=\begin{bmatrix}\cos{\lambda} &\sin{\lambda} &0\\ -\sin{\lambda} &\cos{\lambda} &0\\0&0&1\\\end{bmatrix}
    \;\;\;
    \mathcal{R}^{y'}_\delta=\begin{bmatrix}\cos{\delta} &0 &-\sin{\delta}\\ 0&1&0\\\sin{\delta} &0&\cos{\delta}\\\end{bmatrix}
    \;\;\;
    \mathcal{R}^{z''}_{\phi_o}=\begin{bmatrix}\cos{\phi_o} &\sin{\phi_o} &0\\ -\sin{\phi_o} &\cos{\phi_o} &0\\0&0&1\\\end{bmatrix}
\end{equation}

The goal remains to obtain the expressions in equations \ref{eq:expandedKN_Pol} and \ref{eq:expandedKN_Unp} in terms of the detector frame coordinates, i.e. $\mathcal{D}(\vartheta,\varphi)$. So we need to find the transformation of the relevant terms: 
\begin{equation}\label{eq:rotationEqns}
\begin{split}
    \sin^2{\theta}\cos^2{\phi}&=x^2_\gamma\\
    \sin^2{\theta}&=x^2_\gamma+y^2_\gamma\\
    \cos{\theta}&=-z_\gamma\\
\end{split}
\end{equation}
The aforementioned terms are to be transformed according to equation \ref{eq:rotatingToPhotonFrame} to be expressed in terms of the detector frame coordinate system:  $\vartheta,\varphi;\delta,\lambda,\phi_o$. The transformed terms are then substituted in the KN cross-section in equations \ref{eq:expandedKN_Pol} and \ref{eq:expandedKN_Unp} yielding the expected transformed cross-section. To calculate the transformed normalized angular distribution, one would need to evaluate the transformed version of the integrals in \ref{eq:integrals} such that:
\begin{equation}\label{eq:integralsTransformed}
\begin{split}
    \Phi^{\text{pol}}(\varphi;\varepsilon,\delta,\lambda,\phi_o) &= \frac{\int_0^{\pi}\mathcal{D}^{\text{pol}}(\vartheta,\varphi;\varepsilon, \delta,\lambda,\phi_o)\sin{\vartheta} d\vartheta}{\int_0^{2\pi}\int_0^{\pi}{\mathcal{D}^{\text{pol}}(\vartheta,\varphi;\varepsilon, \delta,\lambda,\phi_o)\sin{\vartheta} d\vartheta d\varphi}}\\    \Phi^{\text{unp}}(\varphi,\varepsilon,\delta,\lambda,\phi_o) &= \frac{\int_0^{\pi}\mathcal{D}^{\text{unp}}(\vartheta,\varphi;\varepsilon, \delta,\lambda,\phi_o)\sin{\vartheta} d\vartheta}{\int_0^{2\pi}\int_0^{\pi}{\mathcal{D}^{\text{unp}}(\vartheta,\varphi;\varepsilon, \delta,\lambda,\phi_o)\sin{\vartheta} d\vartheta d\varphi}}
\end{split}
\end{equation}
\edit1{The above-mentioned} equations in \ref{eq:integralsTransformed} would produce the expected modulation curves/distribution for polarized and unpolarized distributions in terms of the measured azimuthal angles in the detector frame $\varphi$ for any source parameters $(\varepsilon,\delta,\lambda,\phi_o)$. We evaluate these integrals \edit1{numerically; the} first-order analytical approximation \edit1{of \cite{fabio}, valid at low energies, is recorded in Appendix \ref{app:lowE} and compared with the numerical result in Section \ref{sec:VerifySimIdeal}}. 

It is important to note here that, these distributions represent the physical aspect/bias of the ASAD without any consideration for the instrumental effects/biases. Some realistic instrumental consideration affect the physical bias of the ASAD as will be covered later. However, it remains true that this approach thus far is blind to the instrumental effects.

\subsection{Generalized Framework}\label{subsubsec:generalCase}
The \edit1{low-energy expansion of Appendix \ref{app:lowE} cannot be carried to the energies of GRB prompt emission, where its expansion parameter is no longer small. No closed form replaces it there,} because the transformed Klein-Nishina integrands in Equation \ref{eq:integralsTransformed} contain products of trigonometric functions of $(\vartheta, \varphi)$ coupled through the rotation matrix elements $\mathcal{R}_{ij}(\delta,\lambda,\phi_o)$. In this work we \edit1{therefore evaluate} the integrals numerically at discretized points $\varphi_i$ \edit1{with no approximation of the cross-section. Throughout, we call this numerically evaluated description the analytical framework, as opposed to the simulated response of an instrument.}

We evaluate the inner integral over $\vartheta$ using fixed-order Gauss-Legendre quadrature\edit1{, whose roots} and weights are computed once \citep{Virtanen2020SciPy} and reused for every evaluation \edit1{(32 nodes on each interval of the analytic coincidence window of Equation \ref{eq:CoincidenceCond} with no gate or with the sharp gate of Equation \ref{eq:stepFn}, and 64 nodes on each of four equal panels of $[0,\pi]$ with the error-function gate of Equation \ref{eq:erfGate})}. The outer integral over $\varphi$ is performed on a fixed grid (360 points over \edit1{$[0,2\pi)$)}, and the normalization in the denominator is \edit1{the periodic mean of the} same grid. The integrands are evaluated \edit1{as vectorized array operations in NumPy \citep{Harris2020NumPy}}. With this implementation, a full ASAD for a given set of beam parameters $(\varepsilon,\delta,\lambda,\phi_o)$ is computed in \edit1{about a millisecond} on a modern workstation \edit1{(the 9,150 mode 1 ASADs of a POLAR response in 11 s),} which makes large parameter sweeps practical.

This approach handles the full parameter space at once, unlike the \edit1{low-energy expansion of Appendix \ref{app:lowE}, which} is limited to the case where $\lambda=\phi_o=0$. Even though the source azimuthal angle $\lambda$ produces a pure phase shift, as shown in Equation 24 of \cite{fabio}, the polarization angle $\phi_o$ changes the shape of the ASAD in a non-trivial, energy-dependent way. 

\begin{figure}[ht!]
\centering
\begin{interactive}{js}{fig06int.zip}
\includegraphics[width=0.82\textwidth]{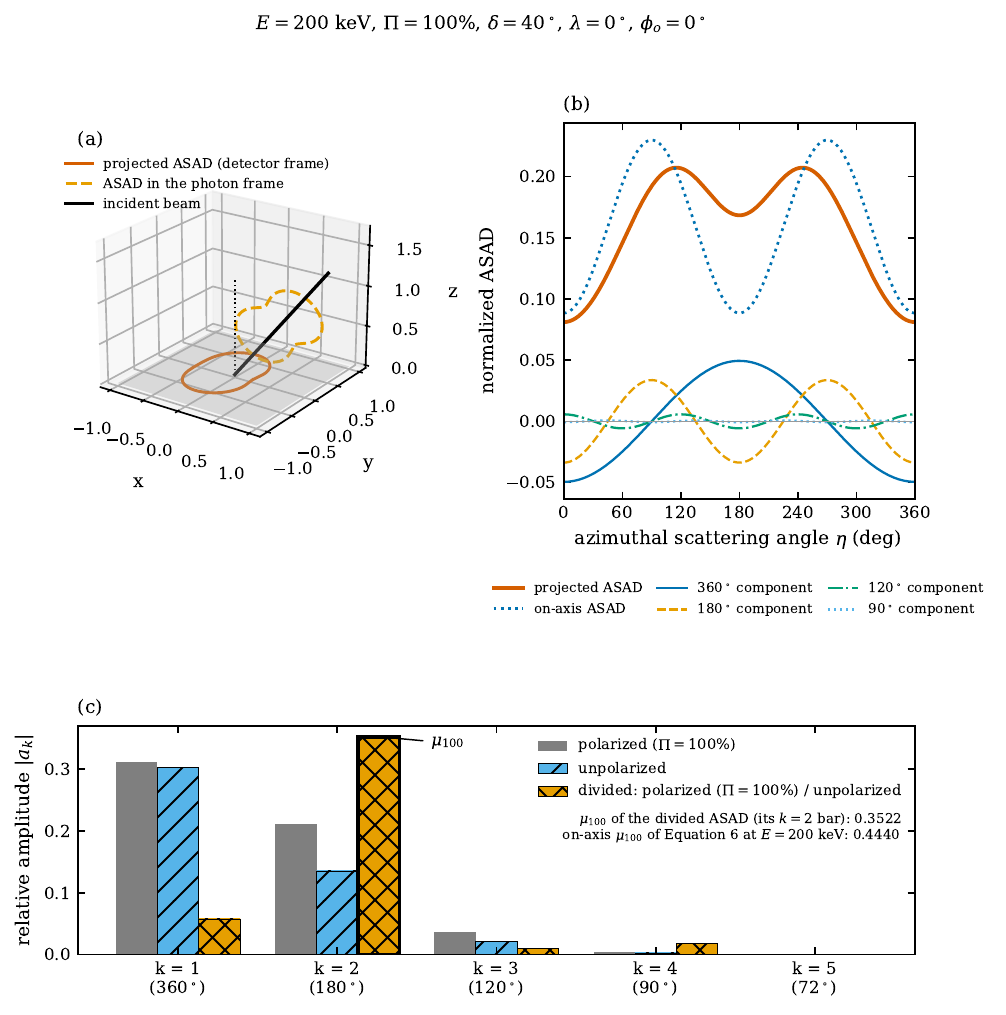}
\end{interactive}
\caption{\edit1{The projected azimuthal scattering angle distribution of an off-axis beam, at $E=200$ keV, $\Pi=100\%$, $\delta=40^\circ$, $\lambda=0^\circ$, $\phi_o=0^\circ$. (a) The ASAD in the photon frame, where the polarization vector defines the $x_\gamma$ axis, drawn in the detector frame by the rotation of Section \ref{subsec:coordTrans}, and the projected ASAD on the detector plane, with the incident beam direction. (b) The projected ASAD compared with the on-axis ASAD at the same beam, and its harmonic components of periods $360^\circ$, $180^\circ$, $120^\circ$ and $90^\circ$. (c) The relative harmonic amplitudes $|a_k|$, $k=1$ to $5$ of the fully polarized projected ASAD, of the unpolarized one and of their ratio, the divided (corrected) ASAD, matching the three columns of Figure \ref{fig:harmonicsAmplitudes}. The outlined $k=2$ bar for the divided ASAD is the $\mu_{100}$ of the classical analysis of Section \ref{sec:classical}. The interactive version recomputes every panel from the transformed Klein-Nishina cross-section as the polarization degree, energy, zenith angle, azimuth and polarization angle are varied. It can also switch on the threshold gate of Equation \ref{eq:erfGate} with adjustable $\mu$ and $\sigma$.}}
\label{fig:visualizeProjections}
\end{figure}

\subsection{Verification Setup}\label{subsec:verifySetup}

\begin{figure}[ht!]
\centering
\includegraphics[width=0.6\textwidth]{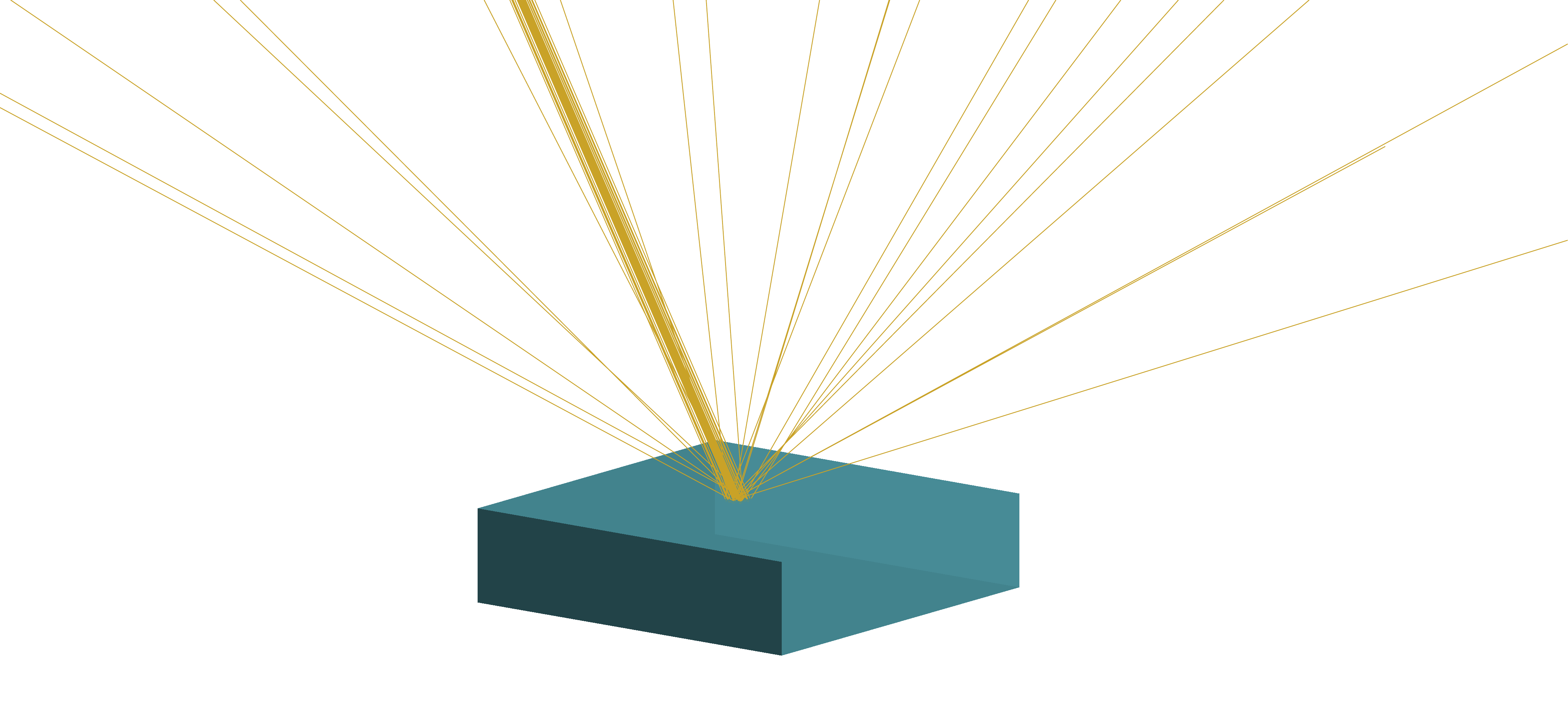}
\caption{The geometry of the ideal \edit1{GEANT4} simulation: a continuous slab made entirely of hydrogen with high density irradiated by a plane-parallel mono-energetic beam incident with zenith angle $\delta=40^\circ$. The scattered photon momentum $\vec{k'}$ is recorded immediately after the first Compton scatter, and used to calculate the azimuthal scattering angles $\varphi$ in the detector plane.}
\label{fig:idealPolarimeter}
\end{figure}

To verify the results of our framework, we utilized a physics-based \edit1{GEANT4} simulation to produce the ideal detector response \citep{GEANT4_2003}. A response that captures purely the physics of the scattering process, without instrumental effects as expected by KN cross-section when projected onto a two-dimensional planar detector as in Figure \ref{fig:visualizeProjections}. We set up a dense slab made entirely of hydrogen acting as a perfect scatterer, with a density of $100\,\mathrm{g/cm^3}$ to maximize the probability of Compton interaction while keeping essentially all interactions within the top layer of the artificial scatterer. We then fired a plane parallel beam of mono-energetic photons at the detector. A schematic of the simulation geometry is shown in Figure \ref{fig:idealPolarimeter}. 

We were able to fire photons from any source direction denoted by zenith $\delta$ and azimuth $\lambda$, polarized with any polarization angle $\phi_o$ or unpolarized. We used GEANT4's \texttt{GetPostStepPoint()} to record the photon's scattering direction $\vec{k'}$ after its first interaction in the detector. Using $\vec{k'}$ and the predefined $xy-$axes of the detector, we were able to determine the azimuthal scattering angle $\varphi$ whose histogram constitutes the ASAD. We filtered out photons that underwent photoelectric absorption as photons had \texttt{NaN} energy after scattering due to being fully absorbed. We also filtered out photons that had no change in energy after interaction as they constitute Rayleigh scattering. We also minimized Rayleigh scattering cross-section by choosing Hydrogen for our material since it is roughly proportional to $Z^2$, \citep{attix1986}. With these selections, the produced ASADs purely captured the physics of Klein-Nishina cross-section as anticipated without any additional instrumental or physical effects. We generated ASADs for multiple beam parameters to compare with the results of our numerical calculations. The results of these comparisons are presented in Section \ref{VerifySim}.

\section{Realistic Instrument Considerations}\label{sec:realInstrument}
\subsection{Energy Thresholds and Kinematic Limits}\label{sec:threshold}
\begin{figure}[ht!]
\centering
\includegraphics[width=0.7\textwidth]{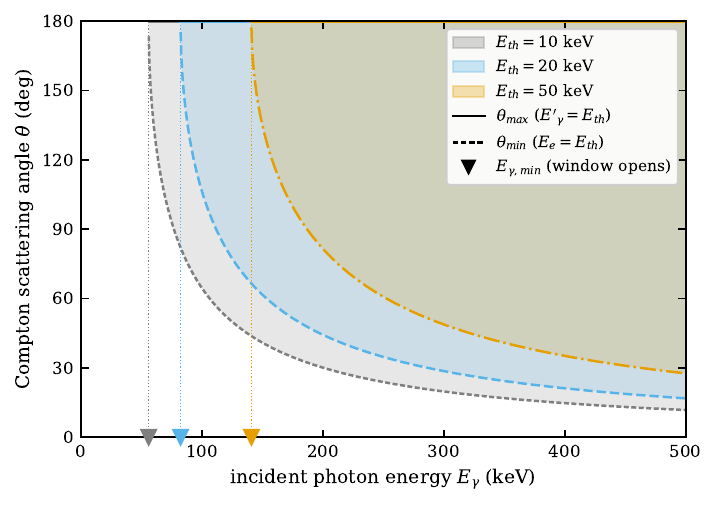}
\caption{Valid Compton coincidence window as a function of incident photon energy $E_\gamma$ for three detector energy thresholds: $E_{th} = 10, 20 \;\text{and}\; 50 \;\text{keV}$. For each threshold, the shaded band spans the range of scattering angles $\theta_\text{min}<\theta<\theta_\text{max}$ for which both the recoil electron and the scattered photon deposit energy above $E_{th}$, satisfying the coincidence condition. The solid upper boundary $\theta_{\text{max}}$ is set by the scattered photon condition $E'_\gamma=E_{th}$
(Equation \ref{eq:CoincidenceCond}, left), and the dashed lower boundary $\theta_\text{min}$ is set by the recoil electron condition $E_e=E_{th}$ (Equation \ref{eq:CoincidenceCond}, right). For the two lower thresholds, the scattered photon condition is trivially satisfied across the plotted energy range and $\theta_\text{max}$ remains pinned at $180^{\circ}$. The downward triangle on the $x-$axis marks the crossover energy $E_{\gamma, \text{min}}$ for each threshold, below which no scattering angle can simultaneously satisfy both conditions and the window for coincident detection closes entirely. The window widens with increasing $E_\gamma$ and shrinks with increasing $E_\text{th}$, directly reducing the fraction of the \edit1{Klein-Nishina} angular distribution accessible to the instrument and therefore limiting its polarimetric sensitivity.}
\label{fig:VisualizeComptonCoincidence}
\end{figure}

The integrals in equation \ref{eq:integralsTransformed} assume a full range of Compton scattering angles can be measured with a detector, but in general this range will be affected by the pre-set energy detection thresholds in the instrument $E_{\text{th}}$\footnote{Strictly, the effective threshold is the sum of the detector channel threshold and the electron binding energy, since the recoil electron is not free. For the low-$Z$ scintillators considered here the binding energy is small and is absorbed into the effective $E_{\text{th}}$.}. Polarimetry events ideally consist of photons that Compton-scatter off a detector element and are subsequently fully absorbed in a second detector element, typically referred to as two-hit or Compton events. Both detector elements must register energy depositions above the instrument's trigger threshold for the event to be identified as a Compton coincidence. A photon with a total energy $E_\gamma$ will deposit energy to the recoil electron $E_e=E_\gamma-E_\gamma'$ and deposit its remaining energy $E'_\gamma$ in the second detector element. Thus, the conditions for a detector to record Compton events are $E_e>E_{\text{th}}$ and $E'_\gamma>E_{\text{th}}$. Figure \ref{fig:VisualizeComptonCoincidence} shows the range of valid Compton scattering angles $\theta_\text{min}<\theta<\theta_\text{max}$ as a function of incident photon energy for three representative threshold energies, illustrating how the window for coincident detection narrows and closes entirely below a threshold-dependent crossover energy $E_{\gamma{,\text{min}}}$. Using the Compton kinematics equation \ref{compton_eq}, we can express the scattering angles as:  
\begin{equation}
    \cos{\theta}=1-\frac{1}{\varepsilon}\Bigg(\frac{E_\gamma}{E_\gamma'}-1\Bigg)    
\end{equation}
The condition on the scattered photon energy sets an upper limit on the scattering angle, $\theta_{max}$, since larger angles correspond to greater energy transfer to the recoil electron, leaving the scattered photon with insufficient energy to trigger the second detector element. Conversely, the condition on the recoil electron energy sets a lower limit $\theta_{min}$, since smaller angles deposit less energy into the first detector element, i.e., leaving a lower energy recoil electron. The boundary angles are found by setting each condition to equality:
\begin{equation}\label{eq:CoincidenceCond}
\begin{split}
    E_\gamma'&\geq E_{\text{th}}\\
    E_\gamma'&=E_{\text{th}}\\
    \cos{\theta_{\text{max}}}&=1-\frac{1}{\varepsilon}\Bigg(\frac{E_\gamma}{E_\text{th}}-1\Bigg)\;\;\;\;\;\;\;\;\\   
\end{split}
\begin{split}
E_\gamma-E_\gamma'&\geq E_{\text{th}}\\ 
E_\gamma'&=E_\gamma-E_{\text{th}}\\ 
\cos{\theta_{\text{min}}}&=1-\frac{1}{\varepsilon}\Bigg(\frac{E_\text{th}}{E_\gamma-E_\text{th}}\Bigg) \\
\end{split}
\end{equation}

While the angular limits $\theta_\text{min}<\theta<\theta_\text{max}$ are straightforward to compute in the photon frame, translating them into limits on the transformed angle $\vartheta$ in the detector frame is non-trivial. From Equation \ref{eq:rotationEqns} where $\cos{\theta}=-z_\gamma(\vartheta,\varphi)$ so that: 
\begin{equation}
    \cos{\theta}=-z_\gamma(\vartheta,\varphi)=-\big(\mathcal{R}_{31}\sin{\vartheta}\cos{\varphi}+\mathcal{R}_{32}\sin{\vartheta}\sin{\varphi}\edit1{-}\mathcal{R}_{33}\cos{\vartheta}\big)
\end{equation}
This expression cannot be easily inverted to give $\vartheta$ as a function of $\varphi$ alone, since the integral limits would then vary at every point along the $\varphi$ axis, so the adaptive quadrature routine would need to solve a transcendental equation for each $\varphi_i$ to find $\vartheta$ limits, making numerical evaluation impractically slow. A practical solution is to instead introduce a Heaviside gate function: 
\begin{equation}\label{eq:stepFn}
    \mathcal{H}(\vartheta,\varphi)=\begin{cases}
    1 \;\;\;\;&\text{if}\;\;\;\;\;\;\;\;\cos{\theta_{\text{max}}}\leq -z_\gamma(\vartheta,\varphi)\leq \cos{\theta_{\text{min}}}\\
    0 &\text{otherwise}
    \end{cases}
\end{equation}
This function gates the integrand, contributing unity only for physically valid scattering geometries and zero otherwise, without altering the integration limits. The integrals in Equation~\ref{eq:integralsTransformed} are then replaced by\footnote{For visual clarity, we suppressed the dependencies on the parameters $\varepsilon,\delta, \lambda,\phi_o$ and $E_\text{th}$}: 
\begin{equation}\label{eq:integralsTransformedStepFns}
    \Phi^{\text{pol}}(\varphi) = \frac{\int_0^{\pi}\mathcal{H}(\vartheta,\varphi)\mathcal{D}^{\text{pol}}(\vartheta,\;\varphi)\sin{\vartheta} d\vartheta}{\int_0^{2\pi}\int_0^{\pi}\mathcal{H}(\vartheta,\varphi){\mathcal{D}^{\text{pol}}(\vartheta,\;\varphi)\sin{\vartheta} d\vartheta d\varphi}}
    \;\;\;\;\;\;\;\;
    \Phi^{\text{unp}}(\varphi) = \frac{\int_0^{\pi}\mathcal{H}(\vartheta,\varphi)\mathcal{D}^{\text{unp}}(\vartheta)\sin{\vartheta} d\vartheta}{\int_0^{2\pi}\int_0^{\pi}\mathcal{H}(\vartheta,\varphi){\mathcal{D}^{\text{unp}}(\vartheta)\sin{\vartheta} d\vartheta d\varphi}}
\end{equation}
In the aforementioned equation, it is shown that the solid angle in the photon frame $d\Omega_\gamma=\sin{\theta}d\theta d\phi$ is equivalent to the transformed solid angle in the instrument frame $d\Omega=\sin{\vartheta}d\vartheta d\varphi$, since the Jacobian of the rotation matrices is $|\mathcal{J}|=1$.

\begin{figure}[ht!]
\centering
\includegraphics[width=0.65\textwidth]{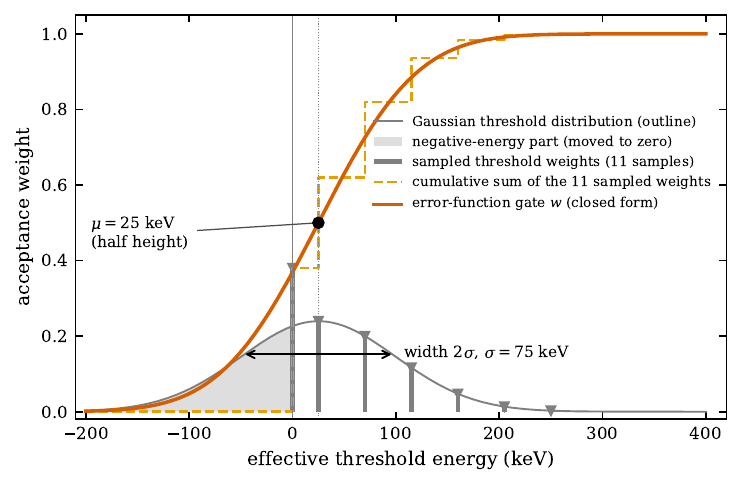}
\caption{\edit1{The threshold gate for the POLAR values $\mu=25$ keV and $\sigma=75$ keV, three elements on one axis. The stems are the sampled threshold weights of the construction of Appendix \ref{app:gate}: 11 thresholds sampled at equal spacing over $\mu\pm3\sigma$, each with its Gaussian weight, the samples at negative energy combined at zero; they sum to one. The staircase is the cumulative sum of those weights, which is the acceptance of the sampled construction. The thin outline is the Gaussian threshold distribution over the full range of effective threshold energies; its part at negative energies (shaded) is the mass that is moved to zero. The solid curve is the closed-form limit, the error-function gate of Equation \ref{eq:erfGate}, read at a deposit equal to the effective threshold energy on the axis, that is against the smaller of the two energy deposits, with the mean $\mu$ marked as the half-height point and $\sigma$ as the width of the turn-on.}}
\label{fig:gate}
\end{figure}

In a real instrument the detection efficiency does not switch on sharply at \edit1{a single threshold energy}. The trigger efficiency as a function of deposited energy instead follows a gradual turn-on\edit1{, softened by} the combined action of finite energy resolution, channel-to-channel gain \edit1{and threshold} variations, electronic noise, \edit1{optical crosstalk between detector elements, and the event selection. We describe the turn-on by a Gaussian distribution of effective thresholds with mean $\mu$ and width $\sigma$. Averaging} the sharp gate of Equation \edit1{\ref{eq:stepFn} over that distribution has a closed form (Appendix \ref{app:gate}): because a coincidence survives a threshold exactly when the smaller of the two deposits exceeds it, the Heaviside gate $\mathcal{H}$ is replaced inside the $\vartheta$ integrand by the smooth acceptance weight}
\begin{equation}\edit1{\label{eq:erfGate}
    w(\vartheta,\varphi)=\Phi_{\mathcal{N}}\!\left(\frac{\min\!\left(E_e,\;E'_\gamma\right)-\mu}{\sigma}\right)}
\end{equation}
\edit1{where $\Phi_{\mathcal{N}}$ is the standard normal cumulative distribution function, giving an error-function turn-on; $E_e$ and $E'_\gamma$ are evaluated at $(\vartheta,\varphi)$ through the scattering kinematics, $\mu$ is the center of the effective threshold distribution, and} $\sigma$ \edit1{is its width. Figure \ref{fig:gate} shows the weight for the POLAR values adopted below. This continuum gate is continuous in incident energy by construction and is the threshold model used throughout this work; the sampled mixture of sharp gates from which it is derived remains available in the released code as an option (Appendix \ref{app:gate}). The gate has the form the hardware exhibits: the in-orbit trigger turn-on of each POLAR channel is itself measured to be an error function \citep{Li2018POLARInOrbit}.}

\edit1{For POLAR we adopt $\mu = 25$ keV from the measured average detection threshold, about 200 ADC at the typical in-orbit gain of about 8 ADC per keV \citep{Li2018POLARInOrbit},} and \edit1{$\sigma = 75$ keV as an effective width absorbing the optical crosstalk between neighboring channels, the channel-to-channel gain spread, the energy resolution, and the event selection. A response-level scan of $\sigma$ at fixed $\mu$ against the full polarized simulation grids confirms this choice: the optimum lies inside a flat region containing 75 keV (Section \ref{subsec:sysLim}), while the measured per-channel} turn-on \edit1{width is an order of magnitude smaller, so $\sigma$ is an effective parameter rather than a resolution. For other instruments the prescription is to use the instrument's average detection threshold for $\mu$ and, for $\sigma$, a width representative of the effects that smear the turn-on: for a clean instrument this is simply the energy resolution near the threshold, while for an instrument like POLAR it is an effective parameter to be fitted.} We emphasize that this is deliberately an \emph{effective} single-parameter description, and its limitations\edit1{, including a mild energy dependence of the fitted width,} are discussed in Section \ref{subsec:sysLim}.

\subsection{Geometric and Non-Uniformity Effects}
The physical design and imperfections of a real detector further modify the measured ASAD beyond what is predicted by the KN cross-section. We define a detector's measured ASAD \edit1{$\mathcal{M}(\varphi)$} with the following product:
\begin{equation}
    \mathcal{M}(\varphi)=\mathcal{I}(\varphi)\times\Phi(\varphi)
\end{equation}
Here, $\Phi(\varphi)$ captures the physics-driven distribution and $\mathcal{I}(\varphi)$ encapsulates all instrumental effects. There are two distinct contributions to $\mathcal{I}(\varphi)$: geometric effects and non-uniformity effects. 

Typical polarimeters consist of segmented detectors that introduce a bias in the ASAD like POLAR and POLAR-2 \citep{POLAR-2}. Photons scattered toward adjacent detectors (i.e., at $\varphi=0^\circ,\;90^\circ,\;180^\circ,\;270^\circ$) traverse a shorter path length and are therefore detected with artificially higher probability than photons scattered at intermediate angles. This introduces a $90^\circ$ periodicity to the ASAD. Additionally, multiple scattering within a single detector element before absorption in a second element leads to an incorrectly reconstructed azimuthal angle $\varphi$. A further realistic effect is incomplete absorption of the scattered photon in the second detector element, which biases the reconstructed energy and, through the kinematic threshold conditions, the recovered scattering angle. More generally, any departure from azimuthal symmetry of the instrument geometry relative to the incident beam direction introduces additional geometric modulations in the ASAD. On the other hand, the variation between the different detector elements such as gain variations and threshold differences among detector elements would produce complex, non-periodic structures in the ASAD that persist throughout the observation. These non-uniform effects further degrade the measured polarization signal \citep{Gill2021}. 

Pointing instruments like PoGOLite \citep{Pearce_2012} and XL-Calibur \citep{Abarr_2021} mitigate geometric effects by rotating during observations, but wide-FOV instruments cannot do so, especially for transient sources such as GRBs. In both cases, Monte Carlo simulations are used to model $\mathcal{I}(\varphi)$. These simulations must be calibrated to reproduce the effects of detector geometry, material properties and detection efficiencies.  

The framework presented here computes only $\Phi(\varphi)$ and Monte Carlo simulations and calibrations remain essential to extract $\mathcal{I}(\varphi)$. Because geometric and efficiency effects are largely independent of the photon's polarization state, we adopt the approximation: 
\begin{equation}\label{eq:assump}
    \mathcal{I}^{\text{pol}}(\varphi)\approx\mathcal{I}^{\text{unp}}(\varphi)
\end{equation}
which implies that unpolarized simulations are sufficient to extract $\mathcal{I}(\varphi)$. However, instruments containing passive material surrounding the active detector volume may exhibit polarization-dependent instrumental effects \citep{Kole2025POLAR2}. In that case, this framework would not necessarily hold, depending on the magnitude of the effects, as studied further in Section \ref{sec:fakeGRBResults}. \edit1{Throughout this work we therefore present the response construction in two modes: mode 1, the unpolarized-only mode, which assumes no polarization-dependent instrumental effects, and mode 2, the two-angle mode, which ingests unpolarized simulations like mode 1 and reads the polarization dependence of the instrument from fully polarized simulations at only two polarization angles per energy (Section \ref{sec:sparseBasis}).} The assumption in Equation \ref{eq:assump} \edit1{underlies mode 1 and} allows us to generate polarization responses rapidly as presented in Section \ref{sec:rapid}. In the following section we apply the framework developed here to real polarization analysis.
\section{Application to Polarization Analysis}\label{sec:app}
\subsection{Classical Analysis}\label{sec:classical}
\begin{figure}[ht!]
\centering
\includegraphics[width=0.85\textwidth]{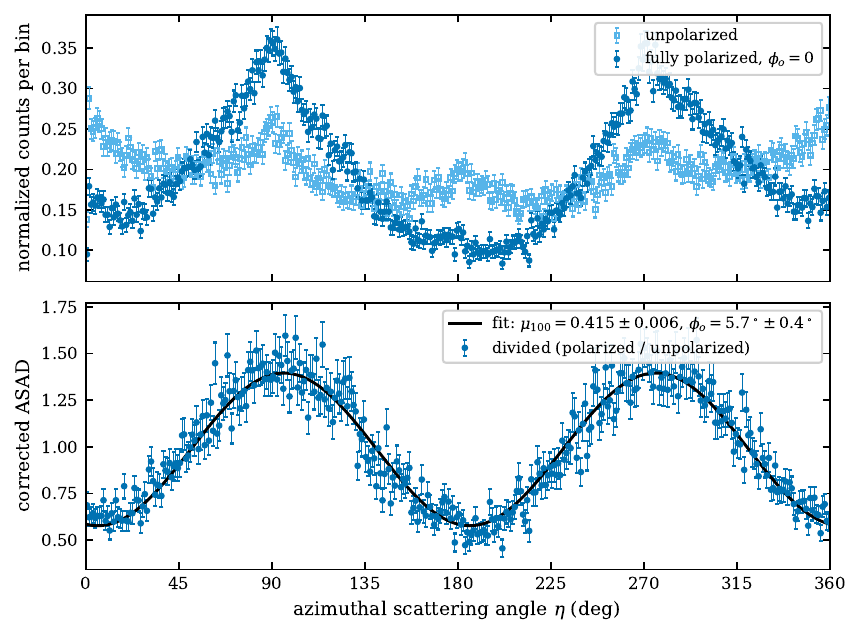}
\caption{An example of the classical analysis approach applied to \edit1{the} fully polarized \edit1{($\phi_o=0$)} and unpolarized ASADs \edit1{at} $E_\gamma= 150$ keV from the POLAR \edit1{response of} GRB 170114A \edit1{(}incident \edit1{at $\delta=27.25^\circ$, Section \ref{sec:VerifySimPOLAR}). Top: the two ASADs} with \edit1{their Poisson errors. Bottom: the divided ASAD, their ratio, with the fit of Equation \ref{onAxisDeriv} giving $\mu_{100}=0.415\pm0.006$ and $\phi_o=5.7^\circ\pm0.4^\circ$. The fitted} angle \edit1{for an input of $\phi_o=0^\circ$ is the phase introduced by the source azimuth $\lambda=6.07^\circ$ of this response (Section \ref{subsubsec:generalCase}), not an error on the measurement}.}
\label{fig:StandardMethod}
\end{figure}

Traditionally, the analysis of polarization data constituted removing geometrical effects and measuring the $\cos^2{\phi}$ modulation in the signal as follows: 
\begin{enumerate}[I]
    \item Histogram the measured azimuthal scattering angle and subtract background to produce the raw uncorrected source ASAD.
    \item Using a measured photon energy spectrum and relative source location the ASAD for both $100\%$ polarized and unpolarized emission are created using Monte-Carlo simulations.
    \item The instrumental component $I(\varphi)$ is removed from the ASAD to produce the corrected ASAD. This is done by dividing the measured ASAD by its unpolarized (simulated) ASAD counterpart, thereby dividing out $I(\varphi)$ and only leaving $\Phi(\varphi)$.
    \item Equation \ref{onAxisDeriv} is then fit to the resulting corrected source ASAD to obtain the fitting parameters $A,\;B,\;\phi_o$. The modulation factor is calculated as \edit1{$\mu=\frac{A}{B}$}.
    \item The value of $\mu_{100}$ can be extracted from the simulated $\Phi(\varphi)$ for this specific measurement. To produce this the simulated polarized and unpolarized ASADs are divided and the result is fitted with equation \ref{onAxisDeriv} producing the normalizing modulation factor: $\mu_{100}$.
    \item The polarization level can now be calculated using $\Pi=\frac{\mu}{\mu_{100}}$ and the polarization angle is $\phi_o$. 
\end{enumerate}
Figure \ref{fig:StandardMethod} shows a simplified example of the classical analysis approach for a single energy fully polarized ASAD to measure $\mu_{100}$ after correcting with an unpolarized ASAD at the same energy. Here we use the simulated mono-energetic POLAR responses \edit1{described in Section \ref{sec:VerifySimPOLAR}}. The correction removes the geometric effects of the instrument, but \edit1{the classical analysis} remains blind to higher harmonics in the signal due to geometric projection of Klein-Nishina cross-section as discussed in this work. For this particular example these harmonics are relatively small, as discussed later in this work.

\subsection{Forward-Folding Analysis(threeML)}\label{sec:3ml}
\begin{figure}[h]
\centering
\includegraphics[width=0.9\textwidth]{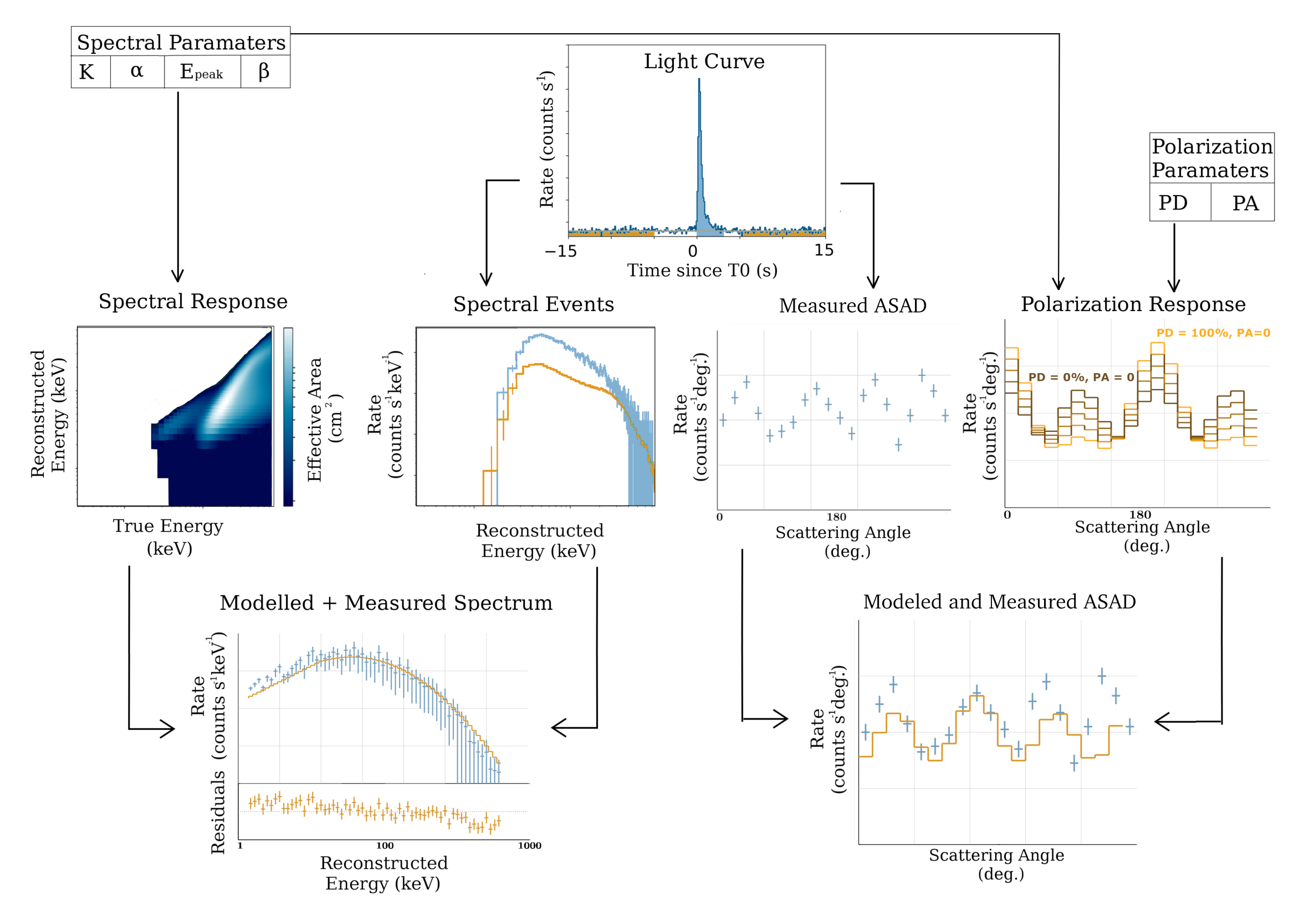}
\caption{A schematic representation of the \edit1{advanced} analysis method applied to GRB analysis. The signal and background interval are selected here to produce a signal and background spectrum and ASAD. The spectral and polarization parameters are folded through a spectral and polarization response to produce modeled ASADs which include the instrumental component $I(\varphi)$. In addition, rather than subtracting the respective background contributions from the data they are added to the modeled spectrum and ASAD allowing these to be directly fitted against the data.}\label{fig:method2}
\end{figure}

The above method is relatively straightforward and requires only two simulations for the analysis, one to produce the $100\%$ polarized ASAD and one for the unpolarized one. A prerequisite of this approach is, however, that the spectral shape of the photons is known, as the shape of both the ASADs depends on this. For common sources like the Crab Nebula this is generally not an issue. However, for a source with an unknown spectrum, such as a Gamma-Ray Burst (GRB), one would have to measure the spectrum first before applying this analysis. To ensure the uncertainty on the spectrum carries on into the polarization properly, one can use the forward-folding approach where a full energy dependent polarization response is produced. In this method, developed in \cite{Burgess2019}, rather than dividing out the $I(\varphi)$ component, it is included in the response and the data can be directly fitted against the response \footnote{This is somewhat analogous to the treatment of background analysis where it can either be subtracted or modeled, the latter is usually preferred to avoid statistical induced effects}. The analysis, also shown schematically in Figure \ref{fig:method2}, typically follows the following steps:
\begin{enumerate}[I]
    \item Generate a full instrument response using Monte-Carlo simulations. This response consists of an array of ASADs simulated for individual energies, polarization angles and polarization levels ($0$ and $100\%$ suffice, as partial polarization is a linear combination of the two).
    \item Similarly, a spectral response is produced using the standard approach used in X-ray and $\gamma$-ray astrophysics. 
    \item Spectral and polarization parameters are picked from a distribution and forward-folded through both the spectral and polarization response. This produces both a modeled ASAD and a modeled spectrum.
    \item The modeled spectrum and the modeled ASAD (which includes $I(\varphi)$ and $\Phi(\varphi)$) can be directly fitted against the data and a combined likelihood can be calculated. 
    \item This method is repeated to produce likelihood distributions for the various spectral and polarization parameters.
\end{enumerate}

This algorithm was originally developed as a plug-in \footnote{\url{https://github.com/grburgess/polarpy}} for the threeML \citep{3ML} framework and could be used to analyze the data from POLAR \citep{Kole2020}. Currently, it is being generalized \edit1{as the \texttt{polpy} package}\footnote{\url{https://github.com/threeML/polpy/}} \edit1{\citep{2026arXiv260827775M}} to be usable for other instruments such as AstroSAT CZTI and the upcoming POLAR-2 and Daksha \edit1{\citep{Bhalerao2024Daksha}} missions. 

The advantage of this approach is that it allows to analyze sources with unknown spectra such as the GRBs observed only by POLAR \citep{Kole2020} for which no spectral information was available from other missions. In addition, through its integration in the threeML framework it allows for joint analysis with data from other instruments which observed the GRB, thereby reducing systematic errors on the spectra which further reduce systematic errors on the polarization. A downside of the method is that it is computationally expensive as it requires a large number of simulations to produce the full response. For POLAR, for each GRB (so for a given incoming angle with respect to the instrument), a 3d grid of ASADs was simulated. This response is computationally expensive to produce and each new GRB detected at a particular location requires a new response to be generated.

\subsection{Rapid Response Generation via Analytical Prediction}\label{sec:rapid}

While the threeML algorithm presented in Section \ref{sec:3ml} is more \edit1{general} than the classical analysis presented in Section \ref{sec:classical} in measuring polarization for off-axis sources simultaneously while also doing spectral fits and joint analysis with other missions, it requires a very large response to be generated. Each response generated to analyze a GRB observed by POLAR took about two weeks on a computer cluster. So for the POLAR catalog of 14 GRBs, it required about 7 months of simulations on a large computer cluster. By utilizing the information presented thus far in this work, a single complete response could be generated much faster as will be shown here. 

We can write the ASAD for an unpolarized and a fully polarized beam for any source with parameters $(\varepsilon, \delta,\lambda,\phi_o)$ as: 
\begin{equation}\label{eq:appSameI}
\begin{split}
        \mathcal{M}^{\text{pol}}(\varphi;\varepsilon,\delta,\lambda)&=\Phi^{\text{pol}}(\varphi;\varepsilon,\delta,\lambda)*\mathcal{I}(\varphi;\varepsilon,\delta,\lambda)\\
        \mathcal{M}^{\text{unp}}(\varphi;\varepsilon,\delta,\lambda)&=\Phi^{\text{unp}}(\varphi;\varepsilon,\delta,\lambda)*\mathcal{I}(\varphi;\varepsilon,\delta,\lambda)
\end{split}
\end{equation}
It was outlined in section \ref{subsubsec:generalCase}, that $\Phi(\varphi)$ can be calculated for any beam parameters $\varepsilon,\delta,\lambda$ and $\mathcal{I}(\varphi)$ can be taken from an unpolarized ASAD under the assumption that there would be negligible polarization-dependent instrumental effects, as follows. \edit1{Starting from} the unpolarized simulated ASAD $\mathcal{M}^{\text{unp}}(\varphi;\varepsilon,\delta,\lambda)$ \edit1{of Equation \ref{eq:appSameI}}, we can peel off $\mathcal{I}(\varphi;\varepsilon,\delta,\lambda)$ by calculating $\Phi^{\text{unp}}(\varphi;\varepsilon,\delta,\lambda)$. Then, we calculate $\Phi^{\text{pol}}(\varphi;\varepsilon,\delta,\lambda)$ which is to be used to calculate a reasonable estimate for the expected simulated fully polarized ASAD $\tilde{\mathcal{M}}^{\text{pol}}_{\phi_o}(\varphi;\varepsilon,\delta,\lambda)$ at a particular polarization angle. That is: 
\begin{equation}\label{eq:appSameIMinusNorm}
    \tilde{\mathcal{M}'}^{\text{pol}}_{\phi_o}(\varphi;\varepsilon,\delta,\lambda) = \frac{\mathcal{M}^{\text{unp}}(\varphi;\varepsilon,\delta,\lambda)}{\Phi^{\text{unp}}(\varphi;\varepsilon,\delta,\lambda)}\times\Phi^{\text{pol}}(\varphi;\varepsilon,\delta,\lambda)\times\frac{\int \Phi^{\text{pol}}(\varphi;\varepsilon,\delta,\lambda) d\varphi}{\int \Phi^{\text{unp}}(\varphi;\varepsilon,\delta,\lambda) d\varphi}
\end{equation}
The last term here is a normalization factor to account for the difference in normalization of $\Phi(\varphi)$ between polarized and unpolarized distributions. We lastly can account for the relative counts in the unpolarized ASAD as follows: 
\begin{equation}\label{eq:appSameIcomplete}
    \tilde{\mathcal{M}}^{\text{pol}}_{\phi_o}(\varphi;\varepsilon,\delta,\lambda) = \tilde{\mathcal{M}'}(\varphi;\varepsilon,\delta,\lambda)\times\frac{\int \mathcal{M}^{\text{unp}} (\varphi;\varepsilon,\delta,\lambda) d\varphi}{\int \tilde{\mathcal{M}'}^{\text{pol}}_{\phi_o}(\varphi;\varepsilon,\delta,\lambda) d\varphi}
\end{equation}
The tilde symbol $\tilde{}$ is meant to mark the expected-calculated distribution $\tilde{\mathcal{M}}^{\text{pol}}_{\phi_o}$, such as that calculated through our approach here, not \edit1{the actual distribution} obtained from a \edit1{GEANT4} simulation $\mathcal{M}^{\text{pol}}_{\phi_o}$. \edit1{This unpolarized-only construction, requiring no polarized simulations, is the response mode we refer to as mode 1 throughout the remainder of this work; by construction it assumes no polarization-dependent instrumental effects (Equation \ref{eq:assump}).} As will be shown later, our calculations for the polarized ASAD matched their counterparts generated by GEANT4 simulations for the POLAR instrument response. We checked responses generated for multiple source positions at different energies and polarization angles. Furthermore, the expected-calculated distribution\edit1{s} can be used to obtain polarization measurements comparable to those obtained using \edit1{GEANT4} polarized simulations. This has been verified for real GRB data as well as brighter \edit1{synthetic GRB data} detected by the POLAR instrument\edit1{, this is} presented in Section \ref{sec:results}.

This approach cuts down the simulation time by a significant factor. A real-life example is the responses typically generated to run polarization measurement analysis using \edit1{the} threeML algorithm outlined in Section \ref{sec:3ml}. For each GRB (specified by its location), a full \edit1{POLAR} response is generated by \edit1{simulating} $150$ \edit1{mono-energetic beams from 5 to 750 keV in 5 keV steps; at each energy the response stores one unpolarized simulation and 61 fully polarized simulations at $3^\circ$ steps from $\phi_o=0^\circ$ to $180^\circ$, where the $180^\circ$ entry repeats the configuration of the $0^\circ$ entry to give the fitting software the closed boundary it expects, so that there are 60 independent polarization angles and 62 simulations per energy, 9,300 simulations in all. Mode 1 replaces the 61 polarized simulations per energy by the analytic construction and needs 1 of the 62 simulations per energy, 98 percent fewer, so the simulation time is} reduced by a factor of \edit1{about 60 (Table \ref{table:costComparison}). Mode 2 (Section \ref{sec:sparseBasis}) needs 3 of the 62, 95 percent fewer, so either mode needs at least 95 percent fewer Monte Carlo simulations than a full response. The polarization-angle step of the generated grid is not tied to the simulated one: both modes refine it in compute only, since the model is analytical.} A single full response was typically produced within 2 weeks on a computer cluster, but with this approach, a full response can be produced within a few hours. 

\subsubsection{Error Propagation on the Analytically Predicted ASAD}\label{sec:errProp}

The analytical distributions $\Phi^\text{pol}(\eta)$ and $\Phi^\text{unp}(\eta)$ are obtained by numerical integration of the Klein-Nishina cross section and carry no statistical uncertainty. The only source of error in the predicted ASAD is therefore the unpolarized simulation $M^\text{unp}(\eta_i)$, whose bin counts follow Poisson statistics. The raw bin counts $M^{\mathrm{unp}}_i\cdot f$ are Poisson-distributed with uncertainty $\sqrt{M^{\mathrm{unp}}_i\cdot f}$; rescaling by $1/f$ to return to the normalized histogram gives the per-bin uncertainty:
\begin{equation}\label{eq:unpErr}
\sigma_{M^\text{unp}_i} = \frac{\sqrt{M^\text{unp}_i \cdot f}}{f}.
\end{equation}

Combining Equations \ref{eq:appSameIMinusNorm} and \ref{eq:appSameIcomplete}, the predicted polarized ASAD can be written as:
\begin{equation}\label{eq:predCombined}
\tilde{\mathcal{M}}^\text{pol}_{\phi_o}(\eta_i) = M^\text{unp}(\eta_i) \cdot \underbrace{\frac{\Phi^\text{pol}(\eta_i)}{\Phi^\text{unp}(\eta_i)}}_{\mathcal{K}_i} \cdot \underbrace{\frac{\int \Phi^\text{pol}(\eta)\, d\eta}{\int \Phi^\text{unp}(\eta)\, d\eta}}_{\mathcal{C}} \cdot \underbrace{\frac{\int M^\text{unp}(\eta)\, d\eta}{\int \tilde{\mathcal{M}'}^\text{pol}_{\phi_o}(\eta)\, d\eta}}_{\mathcal{N}}
\end{equation}
where $\mathcal{K}_i = \Phi^{\mathrm{pol}}(\eta_i)/\Phi^{\mathrm{unp}}(\eta_i)$ is the analytical polarized-to-unpolarized ratio in the $i^{\text{th}}$ bin, $\mathcal{C}$ is a purely analytical normalization constant, and $\mathcal{N}$ is a data-driven normalization that preserves the total number of counts.

To first order, we treat $\mathcal{N}$ as a constant multiplier. This is justified because $\mathcal{N}$ depends on sums over all $N_{\rm bins}$ bins, so the implicit correlation introduced by varying a single $M^\text{unp}_i$ is suppressed by a factor of order $1/N_{\rm bins}$ relative to the direct dependence. Under this approximation, $\tilde{\mathcal{M}}^\text{pol}_i$ is linear in $M^\text{unp}_i$, and the uncertainty propagates as:
\begin{equation}\label{eq:predErr}
\sigma_{\tilde{\mathcal{M}}^\text{pol}_i} = \sigma_{M^\text{unp}_i} \cdot \mathcal{K}_i \cdot \mathcal{C} \cdot \mathcal{N}.
\end{equation}
Equivalently, the fractional uncertainty on the predicted bin equals the fractional uncertainty on the corresponding unpolarized bin:
\begin{equation}\label{eq:predFracErr}
\frac{\sigma_{\tilde{\mathcal{M}}^\text{pol}_i}}{\tilde{\mathcal{M}}^\text{pol}_i} = \frac{\sigma_{M^\text{unp}_i}}{M^\text{unp}_i}.
\end{equation}
We note that this treatment neglects the weak bin-to-bin correlation introduced by the global normalization $\mathcal{N}$. A full covariance treatment would introduce off-diagonal terms of order $1/N_{\rm bins}$ relative to the diagonal, which is negligible for the \edit1{extents} of this work. The uncertainty defined in Equation \ref{eq:predErr} is the $\sigma_{\text{pred},i}$ used in the goodness-of-fit metric of Section \ref{VerifySim}.

\subsubsection{\edit1{The Two-Angle Response Mode (Mode 2)}}\label{sec:sparseBasis}
\edit1{The response mode described above requires no polarized simulations: the polarized ASADs are predicted analytically from the unpolarized prior under the assumption of Equation \ref{eq:assump}. The comparison of that prediction with the full simulated grid of the POLAR responses (Section \ref{sec:VerifySimPOLAR}) motivates a second mode. When the residual of the mode 1 prediction against the simulated polarized ASADs is mapped over energy and polarization angle (Section \ref{sec:VerifySimPOLAR}), it is periodic in the polarization angle with period $180^\circ$. Per azimuthal bin, the simulated polarized ASAD as a function of the polarization angle is therefore described by a constant plus one sinusoid in twice the angle: three numbers. The constant is supplied by the unpolarized simulation, and the two sinusoid coefficients by fully polarized simulations at two angles, we chose $\phi_o=0^\circ$ and $45^\circ$, which isolate the cosine and the sine terms:}
\begin{equation}\edit1{\label{eq:sparseBasis}
    \mathcal{M}^{\text{pol}}_{\phi_o}(\varphi)=\mathcal{M}^{\text{unp}}(\varphi)+\left[\mathcal{M}^{\text{pol}}_{0^\circ}(\varphi)-\mathcal{M}^{\text{unp}}(\varphi)\right]\cos{2\phi_o}+\left[\mathcal{M}^{\text{pol}}_{45^\circ}(\varphi)-\mathcal{M}^{\text{unp}}(\varphi)\right]\sin{2\phi_o}}
\end{equation}
\edit1{That is the whole construction: simulating the unpolarized beam and fully polarized beams at $\phi_o=0^\circ$ and $45^\circ$ for each energy determines the ASAD at every polarization angle, and at every partial polarization fraction, up to the Poisson noise of the input ASADs. Any two polarized simulations at angles whose difference is not a multiple of $90^\circ$ determine the two coefficients (at a multiple of $90^\circ$ the sinusoid repeats up to sign and the pair carries one number); $0^\circ$ and $45^\circ$ isolate the cosine and the sine terms and condition the solution best, and Equation \ref{eq:sparseBasis} is the form for that pair, while for another pair the two coefficients are the solution of a two-by-two linear system per bin, and this is followed for the released code. It is validated on every polarized angle not used as input (Section \ref{sec:VerifySimPOLAR}); as a blind test, every polarized simulation other than the two inputs was removed from the input files and the response rebuilt, and the output was identical. Because the polarized ASADs are simulated rather than predicted, this mode carries the polarization-dependent instrumental effects that mode 1 assumes away, while replacing the 61 polarized simulations per energy of the full response with two. This is the form expected if the instrument response is linear in the polarization state of the beam, as in the on-axis Stokes description of \citet{Kislat2015}; whether that linearity holds exactly off axis and through a real instrument is left to future work. The released code accepts any number of input angles from two upward.}

\begin{deluxetable}{lcccc}
\tablecaption{Cost comparison of the three response-generation approaches for a wide field-of-view Compton polarimeter. \label{table:costComparison}}
\tablehead{
  \colhead{Approach} &
  \colhead{Simulation per energy} &
  \colhead{Simulation wall time} &
  \colhead{Construction} &
  \colhead{$\phi_o$ refinement}
}
\startdata
\edit1{Classical full simulation} & \edit1{61 pol + 1 unp} & \edit1{$\sim$2 weeks per GRB} & \nodata & \edit1{grows with the number of simulated angles} \\
\edit1{Mode 1 (unpolarized only)} & \edit1{1 unp} & \edit1{$\sim$5 hr (est.)} & \edit1{13 s} & \edit1{compute only} \\
\edit1{Mode 2 (two-angle)} & \edit1{1 unp + 2 pol} & \edit1{$\sim$16 hr (est.)} & \edit1{13 s} & \edit1{compute only} \\
\enddata
\tablecomments{Abbreviations: pol = fully polarized simulation, unp = unpolarized simulation, per simulated energy. Wall times refer to the POLAR production grid ($3^\circ$ polarization-angle step, 61 simulations per energy): the classical value is the recorded production time for the per-GRB response on a computing cluster, and the mode 1 and mode 2 values are estimates obtained by scaling that time by the required number of simulated ASADs per energy (about 1/60 and 1/20 respectively), assuming simulation time scales linearly with their number. Construction times are measured on a single workstation for the 61-angle by 150-energy grid of the POLAR responses and will vary with hardware; they include reading the input simulation and writing the response files. Refining the polarization-angle step costs the classical approach simulation time in proportion to the number of simulated angles, while both modes refine it in compute only, within seconds, to arbitrarily fine steps; the polarization fraction is likewise a free, compute-only parameter in both modes.}
\end{deluxetable}

\section{Verification and Systematic Studies}\label{VerifySim}

\subsection{Verification Against Ideal Simulations}\label{sec:VerifySimIdeal}
\begin{figure}[ht!]
\centering
\includegraphics[width=0.9\textwidth]{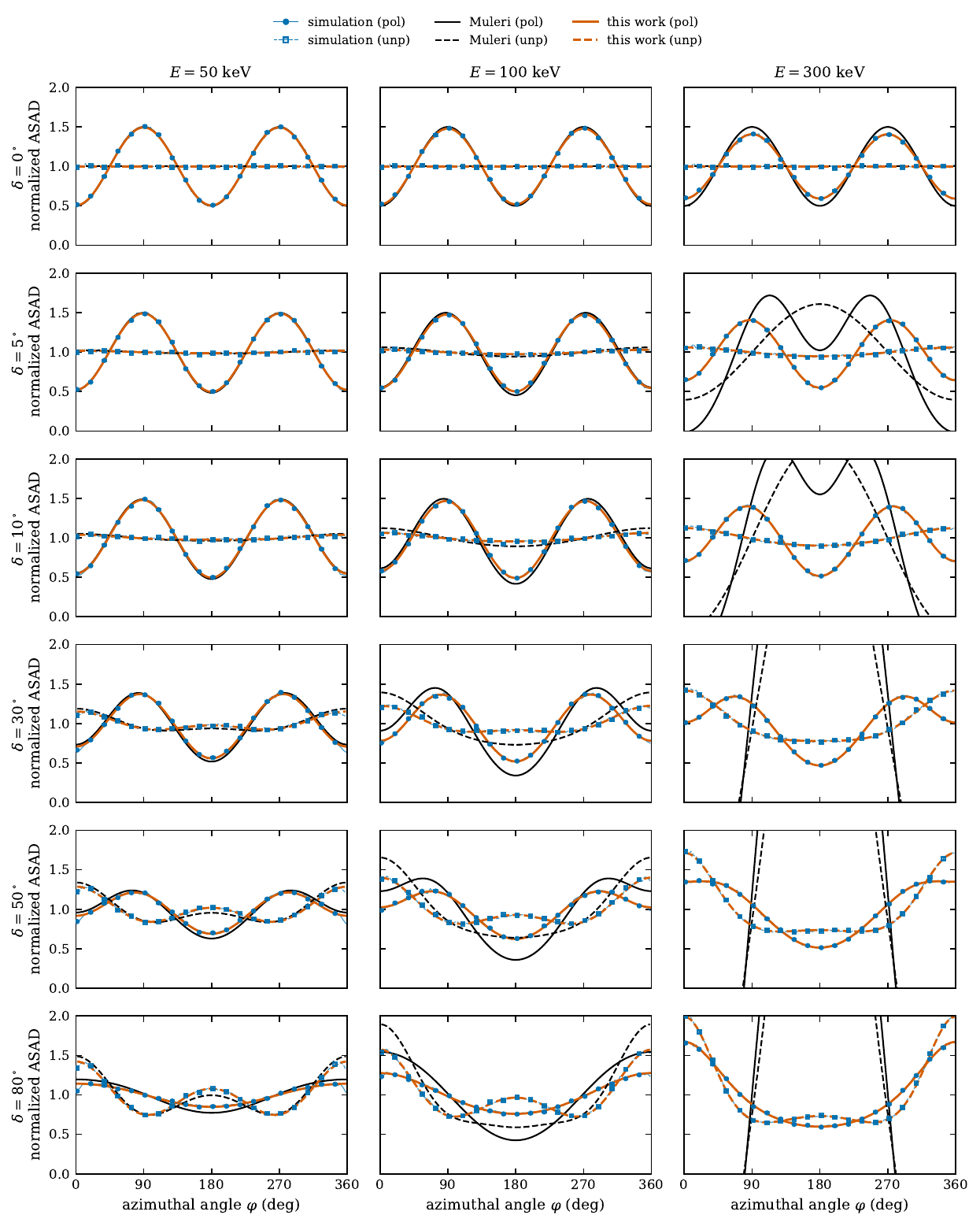}
\caption{Verification of the analytical ASAD framework against the \cite{fabio} low-energy \edit1{expansion (Appendix \ref{app:lowE})} and ideal \edit1{GEANT4} simulations. Each panel shows the normalized azimuthal scattering angle distribution (ASAD) for a fully polarized (solid lines\edit1{, filled markers}) and unpolarized (dashed lines\edit1{, open markers}) mono-energetic beam. Columns correspond to different photon energies, and rows correspond to different source zenith angles. Three curves are shown in each panel: the numerical Klein-Nishina integration presented in this work (\edit1{orange}), the first-order \edit1{expansion} of \cite{fabio} (\edit1{black}), and the ideal \edit1{GEANT4} simulation using a Z = 1 (hydrogen) scattering slab (\edit1{blue}, with Poisson error bars shown every five \edit1{bins}). All distributions are normalized to unit mean.}
\label{fig:VerifySim}
\end{figure}

We compared the ASADs obtained from the ideal detector simulation, as presented in Section \ref{subsec:verifySetup}, to the \edit1{low-energy expansion of \cite{fabio} (Appendix \ref{app:lowE})}, and the analytical ASAD derived in this work for a grid of beam energies and zenith locations. We are showing polarized and unpolarized ASADs for each panel. The results are presented in Figure \ref{fig:VerifySim}. They show that our analytical framework agrees with the simulation \edit1{over the grid of Figure \ref{fig:VerifySim}, 50 to 300 keV and zenith angles to $80^{\circ}$,} while the \cite{fabio} approximation diverges at higher energies where the condition $\varepsilon\ll1$ breaks down, most visibly at 300 keV and large zenith angles. \edit1{The POLAR responses of Section \ref{sec:VerifySimPOLAR} extend the comparison to 750 keV through a full instrument model, and no approximation enters the integrand at higher energies. The framework therefore applies across the band of Section \ref{sec:intro}, with direct verification to 750 keV.}

\subsection{Verification Against POLAR Instrument Response}\label{sec:VerifySimPOLAR}
\edit1{For the case study we} use POLAR responses generated using GEANT4 that utilize the full mass model of the instrument. Each response was created for a specific GRB observation, defined by its incidence angle with respect to the instrument, to be used in its polarization measurement. The responses consist of polarized and unpolarized ASADs for \edit1{the} grid of E and $\phi_o$ \edit1{described in Section \ref{sec:rapid}: 150 energies, one unpolarized simulation and 61 polarized simulations per energy at $3^\circ$ steps from $0^\circ$ to $180^\circ$, of which 60 are independent angles. The POLAR responses and event data used here were obtained from the POLAR collaboration through co-author H.L.} For the purpose of this validation, we selected GRB 170101A representing roughly an on-axis case, GRB 170114A representing an intermediate off-axis case, and GRB 170210A representing an extreme off-axis case. \edit1{The zenith angle of each comparison is taken from the instrument attitude recorded with the data products, and the azimuth from the phase of the simulated polarized distributions, which resolve it to half a degree: GRB 170101A at $\delta=5.76^\circ$, $\lambda=72.86^\circ$; GRB 170114A at $\delta=27.25^\circ$, $\lambda=6.07^\circ$; and GRB 170210A at $\delta=82.59^\circ$, $\lambda=130.9^\circ$.} For all comparisons with the POLAR responses, we used the \edit1{continuum gate of Equation \ref{eq:erfGate} with the adopted POLAR values of Section \ref{sec:threshold}, $\mu = 25$ keV and $\sigma = 75$ keV,} to build each predicted ASAD.

We utilize a modified form of the reduced chi-squared $\chi^2_\nu$ metric as a single summary statistic to assess the agreement between our predicted ASAD and its counterpart from the POLAR response \citep{Gagunashvili2010}. So the per-bin chi-squared $\chi^2_i$ is defined as follows: 
\begin{equation}\label{eq:Chi}
\chi^2_i = \edit1{\frac{\left( y_{\text{true},i} - y_{\text{pred},i} \right)^2}{y_{\text{pred},i}/f + \sigma_{\text{pred},i}^2}}=\frac{\Delta_i^2}{V_i}
\end{equation}
Such that: \edit1{the first term of the denominator is the Poisson variance of the response polarized simulations, evaluated at the model expectation $y_{\text{pred},i}$, with $f$ the normalization factor of Equation \ref{eq:unpErr};} and $\sigma_{\text{pred},i}$ represents the propagated errors of predicted counts $y_{\text{pred},i}$ which stem from Poisson error of each bin counts in the \edit1{response unpolarized simulations}. And equivalently, $\Delta_i$ represents the residual per bin and $V_i$ represents the variance per bin. We adopt the shorthand $\Delta_i$ and $V_i$ for use in the error propagation later-on. \edit1{We consider the first term from the model rather than from the simulation data, since at finite counts a bin that fluctuates low would otherwise be assigned a smaller variance and therefore a larger weight, which biases the statistic downward.}

\edit1{Two conventions follow from this definition and are applied throughout. First, a bin enters the sum only where the variance is defined and positive, which excludes empty bins. Second, quoted map medians are computed above an energy floor of twice the instrument's average per-bar threshold, that is 50 keV for POLAR at the adopted gate, and on one set of cells for both response modes: the 58 polarization angles that are not inputs to the mode 2 construction of Section \ref{sec:sparseBasis}. Below that floor the coincidence requirement leaves almost no accepted counts, the model variance follows the predicted counts toward zero, and the ratio in Equation \ref{eq:Chi} ceases to carry information; the maps are still displayed over the full range, with the sub-floor region marked.}

Then, the reduced chi-squared $\chi^2_\nu$ is calculated using: 
\begin{equation}\label{eq:redChi}
\chi^2_\nu = \frac{1}{\nu} \sum_{i=1}^{\nu} \chi^2_i 
\end{equation}
Where $\nu$ is the \edit1{number of bins entering the sum}. We note here that, this is not a standard reduced chi-squared in the statistical sense because there are no free parameters being fit. We are comparing two fixed distributions, so there \edit1{is} no degrees-of-freedom \edit1{adjustment} for fitted parameters. While it is appropriate for the purpose of gauging agreement, the value $\chi^2_\nu=1$ does not have the usual interpretation of "statistically acceptable fit at $1\sigma$", but it indicates that the residuals between the two distributions are consistent, on average, with the combined uncertainties of both.

The uncertainty on each per-bin $\chi^2_i$ can be obtained by standard error propagation. Since the two distributions ($\text{ASAD}_\text{true},\; \text{ASAD}_\text{pred}$) can be assumed to be statistically independent, the uncertainty on the residual can be derived as $\sigma(\Delta_i) = \sqrt{V_i}$, and the uncertainty on the total variance would be second order  so it can be safely dropped. 
The uncertainty on the $\chi^2_i$ would be dominated by $\sigma(\Delta_i)$ which can be calculated as :
\begin{equation}\label{eq:perBinChiErr}
\sigma^2(\chi^2_i)=\bigg(\frac{\partial}{\partial \Delta_i}(\chi^2_i )\bigg)^2\sigma^2(\Delta_i) = \bigg(\frac{2\Delta_i}{\sqrt{V_i}}\bigg)^2\;\;\longrightarrow\;\;\sigma(\chi^2_i) = \frac{2|\Delta_i|}{\sqrt{V_i}}=2\sqrt{\chi^2_i}
\end{equation}
Lastly, for the reduced chi-squared $\chi^2_\nu$, we can similarly propagate the errors through Equation \ref{eq:redChi} so that: 
\begin{equation}\label{eq:perBinRedChiErr}
\sigma^2(\chi^2_\nu)=\sum_{i=1}^{\nu}\bigg(\frac{\partial}{\partial (\chi_i^2)}\chi_\nu^2\bigg)^2\sigma^2(\chi_i^2)=\frac{4}{\nu}\bigg(\frac{1}{\nu}\sum_{i=1}^{\nu}\chi_i^2\bigg)=\frac{4}{\nu}\chi_\nu^2\;\;\longrightarrow\;\;\sigma(\chi^2_\nu)=2\sqrt{\frac{\chi^2_\nu}{\nu}}
\end{equation}

\begin{figure}[ht!]
\centering
\includegraphics[width=0.85\textwidth]{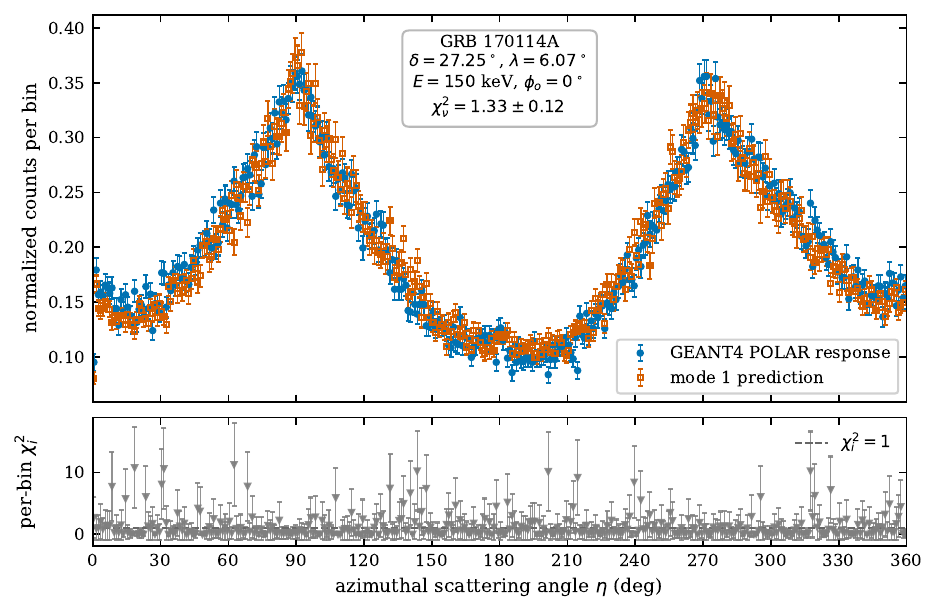}
\caption{A single polarized ASAD comparison between the \edit1{GEANT4-generated} ASAD and the \edit1{mode 1} predicted ASAD at a representative energy $E=150$ keV and polarization angle $\phi_o=0$ for GRB \edit1{170114A,} incident with \edit1{$\delta=27.25^\circ,\; \lambda=6.07^\circ$}. The Poisson errors for the GEANT4 ASAD are shown as well as the propagated errors in the predicted ASAD. The bottom plot shows the per-bin $\chi^2$ values with their respective errors. The overall $\chi^2_\nu$ for these two histograms \edit1{is $\chi_\nu^2=1.33\pm0.12$}.}
\label{fig:singleASAD}
\end{figure}

We show a single ASAD comparison at energy of 150 keV and polarization angle 0 for the \edit1{mode 1} predicted ASAD versus the GEANT4 POLAR response ASAD \edit1{of GRB 170114A} in Figure \ref{fig:singleASAD}. The respective errors per each bin \edit1{are} shown in the Figure and the lower panel represents the per-bin chi-squared $\chi_i^2$ with its associated uncertainties. This \edit1{comparison gives} an overall reduced chi-squared \edit1{$\chi_\nu^2=1.33\pm0.12$ over 360 bins}.

\begin{figure}[ht!]
\centering
\includegraphics[width=0.95\textwidth]{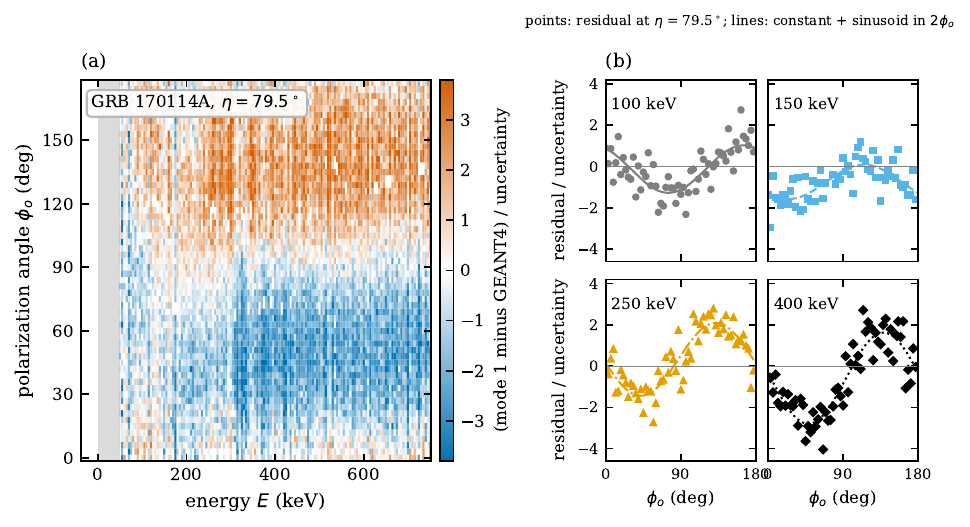}
\caption{\edit1{The mode 1 residual against the GEANT4 polarized ASADs of GRB 170114A, in units of the combined per-bin uncertainty averaged over polarization angle. Left: the residual over energy and polarization angle at the azimuthal bin $\eta=79.5^\circ$, the bin with the largest median sinusoid amplitude above the 50 keV floor (shaded). Right: the residual against polarization angle in the same bin at 100, 150, 250 and 400 keV, one panel per energy on shared axes, each with its points and its fit of a constant plus one sinusoid in $2\phi_o$.}}
\label{fig:mode2Motivation}
\end{figure}

Now, as Figure \ref{fig:singleASAD} shows a decent agreement, we extended the comparisons to \edit1{the full polarized grid of each response, the 150 energies and 60 independent polarization angles of Section \ref{sec:rapid}, for the three GRBs, which span the zenith angle range $\delta\in[5^\circ,83^\circ]$. Before showing the maps of both modes, Figure \ref{fig:mode2Motivation} shows what the mode 1 residual looks like when it is followed across the polarization angle at fixed energy and fixed azimuthal bin for GRB 170114A: the residual is periodic in $\phi_o$ with period $180^\circ$ and is described by a constant plus one sinusoid in $2\phi_o$. At each energy and azimuthal bin, the residual as a function of $\phi_o$ (not of $\eta$) was expanded as a Fourier series in $\phi_o$ up to the sixth term. In each cell, the squared amplitude of the term with a period of $180^\circ$ was divided by the summed squared amplitudes of the six sinusoids. The distribution of that ratio over the 50,695 cells above the 50 keV floor has a median of 91.5 percent and quartiles of 80.1 and 96.1 percent, where a residual with no pattern in $\phi_o$ would spread the sum evenly, one sixth per term on average. This is the observation behind the two-angle construction of Section \ref{sec:sparseBasis}.}

\begin{figure}[ht!]
\centering
\includegraphics[width=0.75\textwidth]{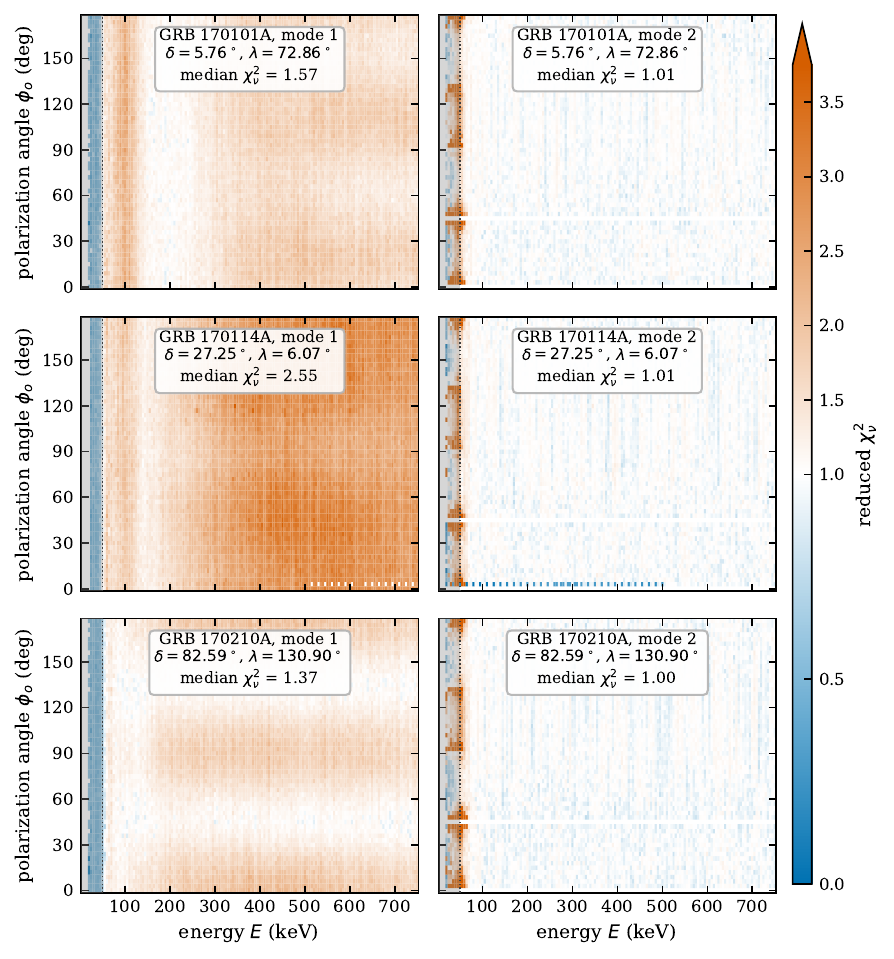}
\caption{\edit1{Reduced $\chi^2_\nu$} of the \edit1{mode 1 (left column) and mode 2 (right column) constructions against the} GEANT4 \edit1{polarized ASADs of GRB 170101A, GRB 170114A} and \edit1{GRB 170210A (rows}, \edit1{by zenith angle), over} the \edit1{150} energies and \edit1{60 independent} polarization angles \edit1{of each response. One color scale is shared by all six panels, white at $\chi^2_\nu=1$, blue below and orange above; the scale tops at the largest quoted mode 1 value, and the few mode 2 cells above it, all at 50 to 65 keV,} are \edit1{drawn at the top color}. \edit1{The region below the 50 keV floor is shaded and does not set the scale. The two input-angle rows of the mode 2 column ($\phi_o=0^\circ$ and $45^\circ$) are blank. Each panel names the comparison geometry and the median over the quoted cells.}}
\label{fig:heatMaps}
\end{figure}

\edit1{Figure \ref{fig:heatMaps} shows the reduced chi-squared maps of both modes for the} three GRBs, \edit1{on one color scale with white at one. Over the quoted cell set the mode 1 medians are 1.57 for GRB 170101A, 2.55 for GRB 170114A and 1.37 for GRB 170210A (75th percentiles 1.74, 2.86 and 1.64), and the mode 2 medians are 1.01, 1.01 and 1.00 (75th percentiles 1.06, 1.06 and 1.06). The mode 1 maps show the evidence for the presumed polarization dependent instrumental effects. The mode 2 maps carry no such structure; the two input-angle rows are blank because they are ingested rather than predicted. A reduced chi-squared below one means that the residuals between the two distributions are smaller than their combined uncertainty, typically, due to low counts}. 

\subsection{Systematic Limitations}\label{subsec:sysLim}
\edit1{The mode 1 construction} provides a physics-based alternative to dedicated polarized simulations, but it \edit1{rests on the approximation of Equation \ref{eq:assump}}. We identify several sources of residual discrepancy \edit1{in the mode 1 construction}: (i) the method's reliance on the unpolarized simulation to capture instrumental effects; (ii) \edit1{real} polarization-dependent instrumental effects, which break the approximation of Equation \ref{eq:assump} and are expected when significant passive material surrounds the detector (for POLAR, the \edit1{entire body of the} Tiangong-2 space station); (iii) residual mismatches between the analytical and simulated ASADs; and (iv) the effective threshold distribution is a single fixed approximation, whereas its physically appropriate width is mildly energy-dependent\edit1{: a response-level scan of $\sigma$ at fixed $\mu$ against the polarized grids of five POLAR responses (the three of this section and those of GRB 170127C at $\delta=41.82^\circ$ and GRB 170305A at $\delta=31.54^\circ$) finds the adopted $\sigma = 75$ keV inside the flat optimum of the full energy band}. We do not attempt to attribute the observed discrepancies to a single cause. As shown in Section~\ref{sec:fakeGRBResults}, the net effect \edit1{of these residuals} is negligible for all but the most extreme, exceptionally bright GRBs\edit1{; the polarization-dependent effects of item (ii), which dominate the mode 1 residual, are modeled directly by the two-angle mode 2, as demonstrated in Section \ref{sec:mode2Results}. However, it is noted that this net effect disappears once mode 2 is employed which reproduces the GEANT4 simulation response down to the statistical noise, therefore there is no impact on the measurement even for extremely bright GRBs.}
\section{Results and Discussions}\label{sec:results}
In this section we validate the \edit1{two response modes of this work} against the full GEANT4 POLAR response in realistic polarization analyses. We first apply both \edit1{modes} to a real POLAR observation\edit1{,} GRB 170114A, whose incidence angle lies in the region where the \edit1{mode 1 construction} is most strained (Section \ref{sec:VerifySimPOLAR})\edit1{, then to} a grid of \edit1{synthetic GRBs with known input polarization parameters that spans the fluence range from a typical bright GRB to the rarest, brightest events, and we close with what the two modes show at the posterior level}.

\subsection{Measuring Polarization for Real GRB Data}\label{sec:resultsRealGRB}
To validate the \edit1{two response modes} against the full GEANT4 response in a realistic polarization measurement, we applied \edit1{all three responses} to the POLAR observation of GRB 170114A (\edit1{$\delta=27.25^\circ$, $\lambda=6.07^\circ$}) using the threeML forward-folding pipeline as described in Section \ref{sec:3ml}. This GRB was selected as the baseline test case for three reasons: its incidence angle coincides with the region where higher order harmonics in the ASAD reach maximum amplitude (Figure \ref{fig:harmonicsAmplitudes}), and the corresponding response exhibits the largest $\chi^2_\nu$ residuals in the comparison between GEANT4 and \edit1{the mode 1} response as shown in Figure \ref{fig:heatMaps}, and it was bright and long enough while exhibiting a polarization measurement with relatively small credibility intervals as described in \cite{Kole2020}.

\begin{figure}[ht!]
\centering
\includegraphics[width=0.75\textwidth]{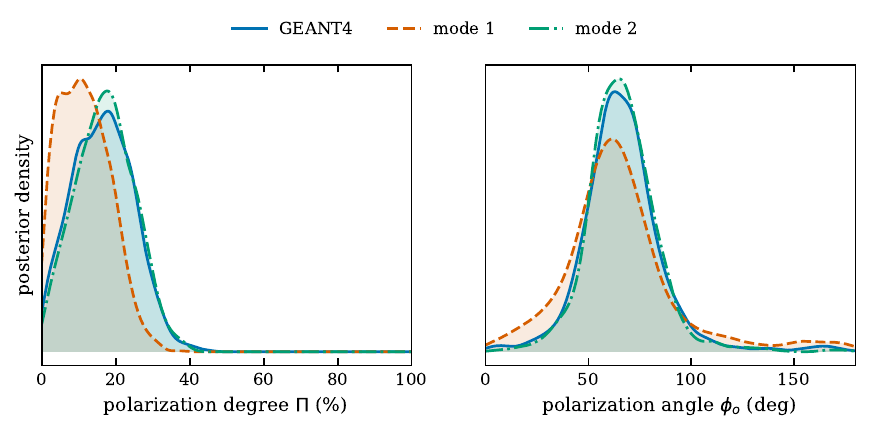}
\caption{\edit1{Posterior distributions of the polarization degree and angle} for \edit1{the} real data \edit1{of GRB 170114A, fitted} with \edit1{the} GEANT4 \edit1{response,} the \edit1{mode 1} response \edit1{and the mode 2 response}.}
\label{fig:realGRB_result}
\end{figure}

The results are shown in the form of posterior distribution\edit1{s} for the polarization parameters $\Pi$ and $\phi_o$ as returned by threeML \citep{3ML} with the \texttt{polpy} plugin for each response. Although \edit1{the analysis fits the} polarization and spectral \edit1{parameters simultaneously, for clarity} we show only the polarization posteriors and omit the corresponding spectral \edit1{ones. The three fits share the identical data, spectral response and sampler configuration; only the polarization response differs. Figure \ref{fig:realGRB_result} shows the result: the GEANT4 response gives $\Pi=16.6^{+8.2}_{-8.8}$ percent and $\phi_o=65.4^{+16.2}_{-14.6}$ degrees, the mode 1 response $\Pi=11.1^{+7.9}_{-6.9}$ percent and $\phi_o=62.5^{+21.1}_{-18.7}$ degrees, and the mode 2 response $\Pi=17.1^{+8.3}_{-8.3}$ percent and $\phi_o=65.7^{+14.9}_{-13.3}$ degrees. At the statistics of this burst the three posteriors overlap, and the mode 2 posterior tracks the GEANT4 one in both parameters}.

\subsubsection{Effect of Increased Prior Statistics}\label{sec:higherStatsResult}
\begin{figure}[ht!]
\centering
\includegraphics[width=0.75\textwidth]{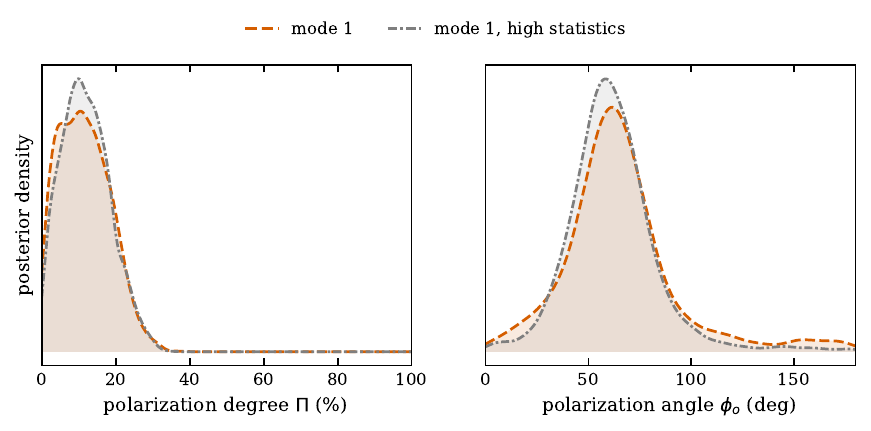}
\caption{\edit1{Posterior distributions} for \edit1{the} real data of GRB 170114A \edit1{with the mode 1} response \edit1{built from the nominal unpolarized prior and with the mode 1} response \edit1{built from the higher-statistics} unpolarized \edit1{prior} (\edit1{30 times the photon count, with its own spectral response}).}
\label{fig:realGRB_result_HigherStatistics}
\end{figure}

Because \edit1{mode 1} relies only on the unpolarized GEANT4 ASADs as a prior, the number of simulations required per response is reduced by a factor of \edit1{about 60.} A natural question is whether redirecting that computational budget into higher-statistics unpolarized simulations would tighten the resulting polarization constraints. To test this, we \edit1{built the mode 1 response of} GRB 170114A \edit1{a second time from a higher-statistics prior: an independent unpolarized simulation of the same burst geometry with its own spectral response, carrying 30 times the photon count of the nominal unpolarized prior ($1.5\times10^{9}$ photons, from the recorded normalization of the simulation).} Figure \ref{fig:realGRB_result_HigherStatistics} shows the resulting posterior distributions \edit1{alongside the nominal mode 1 result of} Figure \ref{fig:realGRB_result}. \edit1{Both responses in this comparison are mode 1 constructions, so polarization-dependent instrumental effects are not accounted for in either. The high-statistics prior gives $\Pi=11.4^{+7.0}_{-6.4}$ percent and $\phi_o=60.0^{+18.0}_{-16.4}$ degrees against $\Pi=11.1^{+7.9}_{-6.9}$ percent and $\phi_o=62.5^{+21.1}_{-18.7}$ degrees for the nominal prior: the credible intervals narrow by 9 percent in degree and 14 percent in angle, and the medians move by +0.3 percentage points and -2.5 degrees, within the intervals. A thirtyfold increase in the prior statistics therefore tightens the measurement only mildly: the width of the posterior is set mainly by the statistics of the data, and the response's own statistical noise is a minor contribution at this brightness}.

\subsection{Diagnosis of Systematics with Simulated GRBs}\label{sec:fakeGRBResults}
The validation in Section \ref{sec:resultsRealGRB} establishes that \edit1{both response modes recover} polarization parameters consistent with the full GEANT4 response for a representative POLAR GRB. However, GRB 170114A is of moderate brightness, and the statistical significance of the measurement is correspondingly limited. To probe the regime in which residual systematics of the \edit1{two response modes} become comparable to the statistical uncertainty, and therefore a dedicated GEANT4 simulation would be required, we constructed a grid of simulated GRBs based on the spectral model and the instrument response for GRB 170114A. Each simulated GRB in the grid was generated with polarization parameters fixed at known input values ($\Pi=10\%$ or $\Pi=50\%$, and $\phi_o=90^\circ$), enabling a direct comparison between measured and true polarization. We adopt the Band-function spectral parameters of GRB 170114A \citep{Kole2020} as the baseline spectrum, and different fluences. The fluence scan is performed at two input polarization degrees: $\Pi=10\%$ representative of the value reported in the POLAR GRB catalog \cite{Kole2020}, and $\Pi=50\%$ which is near the upper end of values predicted by polarization models of GRB prompt-emission and provides a stress test of the method.

\begin{deluxetable}{cccl}
\tablecaption{ Fluence multiples adopted for the simulated GRBs in Section \ref{sec:fakeGRBResults}. The $1\times$ level corresponds to the reported fluence of GRB 170114A \citep{Kole2020}. Approximate all-sky return times are derived from the 10-year Fermi-GBM burst catalog \citep{vonKienlin2020} for long GRBs ($T_{90}>2$s). 
\label{table:fluencesOccur}}
\tablehead{
  \colhead{Fluence multiple} &
  \colhead{Fluence (10--1000 keV)} &
  \colhead{Return time} &
  \colhead{Comparison} \\
  \colhead{} &
  \colhead{(erg\,cm$^{-2}$)} &
  \colhead{(all-sky)} &
  \colhead{}
}
\startdata
$1\times$   & $2 \times 10^{-5}$ & $\sim 1$ week    & GRB 170114A; a typical bright GRB \\
$5\times$   & $1 \times 10^{-4}$ & $\sim 3$ months   & brighter than $\sim98\%$ of GBM detected GRBs \\
$50\times$  & $1 \times 10^{-3}$ & $\sim 3$ years   & $\sim 3$  events per decade of GBM operation \\
$500\times$ & $1 \times 10^{-2}$ & $\sim 100$ years & $\sim 20\times$ fainter than BOAT GRB 221009A \\
\enddata
\end{deluxetable}

\begin{figure*}[ht!]
\centering
\includegraphics[width=0.95\textwidth]{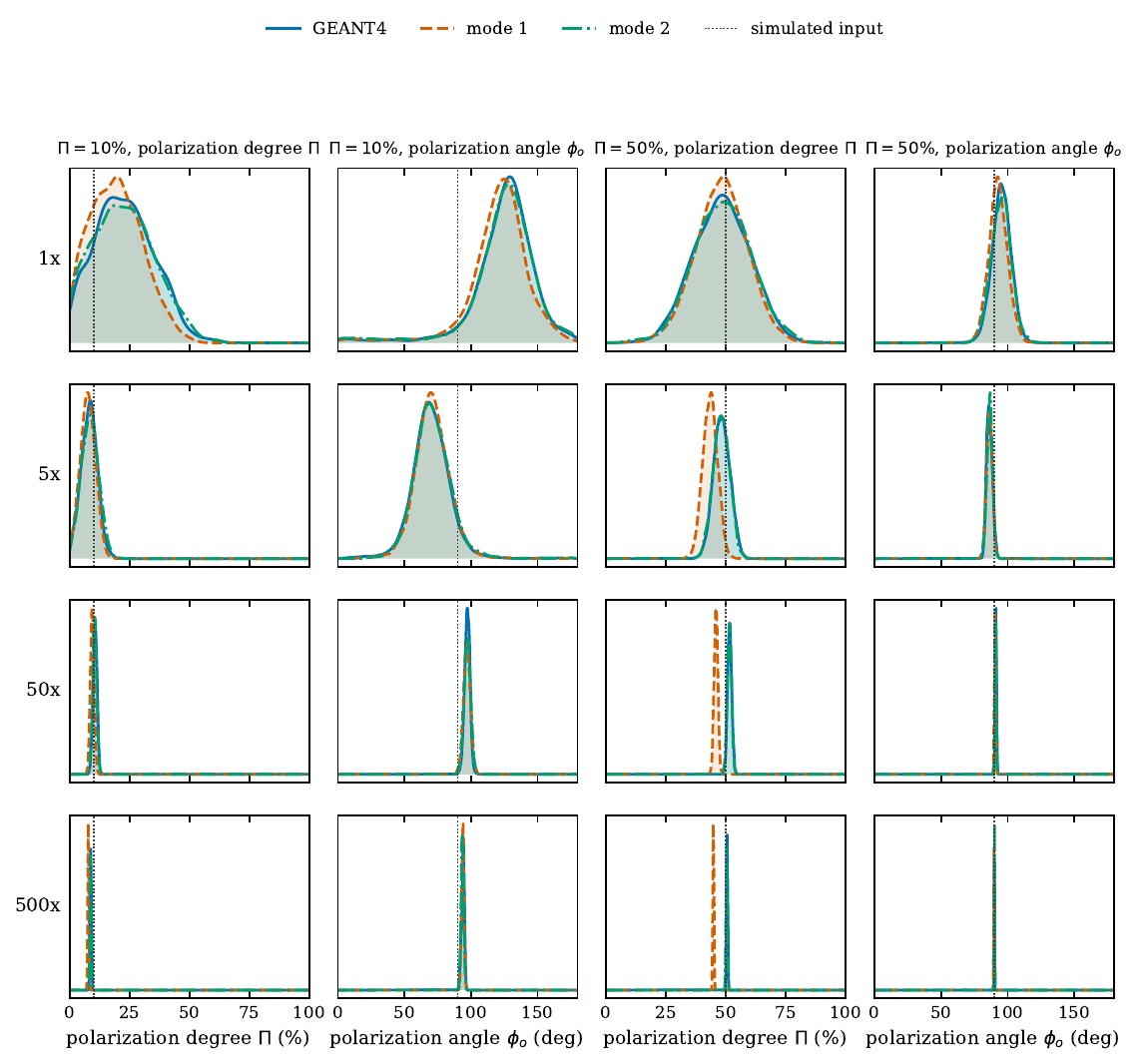}
\caption{Posterior \edit1{distributions of the polarization degree and angle} for \edit1{the} simulated GRB \edit1{grid of} Table \ref{table:fluencesOccur} \edit1{at $\Pi=10\%$ (left two columns) and $\Pi=50\%$ (right two columns), $\phi_o=90^\circ$, fitted with the GEANT4 response, the mode 1 response and the mode 2 response; rows are the fluence multiples, and the dotted lines mark the simulated input values}.}
\label{fig:fakeGRB_result_multipleFluences}
\end{figure*}

GRBs at the fluence level of GRB 170114A ($1\times$) are detected by Fermi-GBM at a rate of roughly once per week, and GRBs at $5\times$ this fluence are detected at a rate of roughly once every 3 months (Table \ref{table:fluencesOccur}). These fluence levels cover the range populated by the vast majority of GRBs a POLAR-class mission will observe during its operational lifetime. At both $\Pi=10\%$ and $\Pi=50\%$, the \edit1{mode 1} and GEANT4 posteriors at $1\times$ and $5\times$ \edit1{are statistically consistent}, as shown in Figure \ref{fig:fakeGRB_result_multipleFluences}\edit1{: the median offsets in $\Pi$ stay within the combined $1\sigma$ credible intervals in every cell except the $\Pi=50\%$, $5\times$ cell, which sits at about 1.0$\sigma$}. In this fluence range, the two responses are statistically indistinguishable in practice, and the \edit1{mode 1} method is therefore sufficient for the overwhelming majority of GRBs observed during a mission lifetime.

For brighter GRBs at $50\times$ and $500\times$ the fluence of GRB 170114A, which are extremely rare, the statistical uncertainty of the recovered $\Pi,\;\phi_o$ collapses significantly, and a small systematic offset between the \edit1{mode 1} and GEANT4 results becomes resolvable. The \edit1{mode 1} response underestimates $\Pi$, indicating that \edit1{the offset} reflects a fixed property of the \edit1{unpolarized-only construction} rather than a statistical fluctuation. We attribute \edit1{it} to the residual polarization-dependent instrumental effects that \edit1{mode 1 assumes away by construction through} Equation \ref{eq:assump}. \edit1{The offset grows} with the degree of linear polarization of the incident beam\edit1{: at $500\times$ the mode 1 median lies 1.0 percentage points below the GEANT4-response median at $\Pi=10\%$ and 5.7 percentage points below it at $\Pi=50\%$ (5.7 at $50\times$). The} recovered polarization angle $\phi_o$ \edit1{is affected more weakly: an angular offset of 0.3$^\circ$ at $\Pi=50\%$ (0.2$^\circ$ at $\Pi=10\%$) becomes resolvable only once the statistical precision collapses at $500\times$, and its absolute scale remains small compared with the credible intervals of GRBs at realistic fluences. This diagnosis defines exactly the regime in which the polarization-dependent effects must be modeled, which is what the mode 2 construction provides.}

\subsection{\edit1{Mode 2 at the Posterior Level}}\label{sec:mode2Results}
\edit1{The mode 2 response of GRB 170114A is built from the unpolarized simulation and the fully polarized simulations at $\phi_o=0^\circ$ and $45^\circ$ at each of the 150 simulated energies, with every other polarized simulation of the grid held out for validation (Section \ref{sec:VerifySimPOLAR}); as a blind test, every polarized simulation other than the two inputs was removed from the input files and the response rebuilt, and the output was identical. We repeated every threeML fit of Sections \ref{sec:resultsRealGRB} and \ref{sec:fakeGRBResults} with this response, using the identical data, spectral response and sampler configuration. Figures \ref{fig:realGRB_result} and \ref{fig:fakeGRB_result_multipleFluences} show the posteriors and Table \ref{table:mode2Posteriors} collects the recovered medians. Mode 2 reproduces the GEANT4-response fit in every dataset within the credible intervals: the polarization-degree medians agree within 0.06 to 0.5 percentage points and the angle medians within 0.7$^\circ$ at the widest and 0.06$^\circ$ at the closest. At the brightest cell ($\Pi=50\%$, $500\times$), where the posteriors are narrowest, the difference is 0.2$^\circ$, at the level set by the statistical noise of the simulated inputs, and the simulated input angle of $90^\circ$ is the angle farthest from both mode 2 inputs. The mode 1 degree offset of Section \ref{sec:fakeGRBResults} is absent in mode 2, which recovers $\Pi=50.3^{+0.3}_{-0.3}$ percent at $\Pi=50\%$ and the highest fluence against the input 50 and the GEANT4-response value of 50.5 percent. At $\Pi=10\%$} the \edit1{GEANT4-response fit itself departs from the simulated input at the brightest fluences: at $500\times$ its polarization degree sits 1.2 percentage points below the input of 10 percent and its angle 4.1$^\circ$ above the input of $90^\circ$, and at $50\times$ the angle sits 7.2$^\circ$ above it, three to four posterior widths in each case; at the brightest fluence the counts in the data reach a tenth of the accepted counts of the simulated response at a typical energy and exceed them at and below 100 keV, so the statistical noise of the simulated response itself may no longer be small against that of the data. Mode 2 reproduces that reference fit rather than the input: the appropriate benchmark for a response construction is the fit obtained with the full simulated response, and mode 2 matches it throughout.}

\begin{deluxetable*}{lcccccc}
\tablecaption{Recovered polarization medians with $1\sigma$ credible intervals for the three responses. Simulated inputs: $\phi_o=90^\circ$ and the stated $\Pi$; the real GRB has no known truth. \label{table:mode2Posteriors}}
\tablehead{
  \colhead{Dataset} &
  \colhead{$\Pi$ GEANT4 [\%]} & \colhead{$\Pi$ mode 1 [\%]} & \colhead{$\Pi$ mode 2 [\%]} &
  \colhead{$\phi_o$ GEANT4 [$^\circ$]} & \colhead{$\phi_o$ mode 1 [$^\circ$]} & \colhead{$\phi_o$ mode 2 [$^\circ$]}
}
\startdata
\edit1{Real 170114A} & \edit1{$16.6^{+8.2}_{-8.8}$} & \edit1{$11.1^{+7.9}_{-6.9}$} & \edit1{$17.1^{+8.3}_{-8.3}$} & \edit1{$65.4^{+16.2}_{-14.6}$} & \edit1{$62.5^{+21.1}_{-18.7}$} & \edit1{$65.7^{+14.9}_{-13.3}$} \\
\edit1{1x, $\Pi = 10\%$} & \edit1{$22.4^{+13.4}_{-11.9}$} & \edit1{$19.1^{+11.3}_{-11.0}$} & \edit1{$22.2^{+13.5}_{-13.2}$} & \edit1{$127.6^{+16.4}_{-18.8}$} & \edit1{$123.1^{+15.5}_{-18.4}$} & \edit1{$127.4^{+16.3}_{-19.2}$} \\
\edit1{5x, $\Pi = 10\%$} & \edit1{$8.3^{+3.3}_{-3.6}$} & \edit1{$7.6^{+3.2}_{-3.2}$} & \edit1{$8.5^{+3.7}_{-3.6}$} & \edit1{$69.5^{+12.7}_{-12.0}$} & \edit1{$70.5^{+12.6}_{-11.5}$} & \edit1{$70.2^{+13.2}_{-12.0}$} \\
\edit1{50x, $\Pi = 10\%$} & \edit1{$10.6^{+0.8}_{-0.8}$} & \edit1{$9.4^{+0.8}_{-0.8}$} & \edit1{$10.5^{+0.8}_{-0.8}$} & \edit1{$97.2^{+1.9}_{-2.0}$} & \edit1{$96.9^{+2.6}_{-2.6}$} & \edit1{$97.3^{+2.4}_{-2.4}$} \\
\edit1{500x, $\Pi = 10\%$} & \edit1{$8.8^{+0.3}_{-0.3}$} & \edit1{$7.8^{+0.2}_{-0.2}$} & \edit1{$8.7^{+0.3}_{-0.3}$} & \edit1{$94.1^{+0.9}_{-1.4}$} & \edit1{$93.8^{+0.8}_{-0.9}$} & \edit1{$93.5^{+1.0}_{-0.9}$} \\
\edit1{1x, $\Pi = 50\%$} & \edit1{$48.2^{+11.9}_{-11.8}$} & \edit1{$48.3^{+10.5}_{-10.8}$} & \edit1{$48.6^{+11.8}_{-11.8}$} & \edit1{$95.4^{+7.0}_{-6.9}$} & \edit1{$92.7^{+7.1}_{-6.5}$} & \edit1{$95.3^{+7.0}_{-7.8}$} \\
\edit1{5x, $\Pi = 50\%$} & \edit1{$48.2^{+3.6}_{-3.4}$} & \edit1{$43.4^{+3.1}_{-3.1}$} & \edit1{$48.0^{+3.4}_{-3.4}$} & \edit1{$86.2^{+2.3}_{-2.0}$} & \edit1{$86.3^{+2.2}_{-2.2}$} & \edit1{$86.7^{+2.0}_{-2.2}$} \\
\edit1{50x, $\Pi = 50\%$} & \edit1{$51.5^{+0.9}_{-0.9}$} & \edit1{$45.8^{+0.7}_{-0.8}$} & \edit1{$51.5^{+0.8}_{-0.9}$} & \edit1{$91.2^{+0.5}_{-0.5}$} & \edit1{$90.6^{+0.5}_{-0.5}$} & \edit1{$91.1^{+0.5}_{-0.5}$} \\
\edit1{500x, $\Pi = 50\%$} & \edit1{$50.5^{+0.3}_{-0.3}$} & \edit1{$44.7^{+0.2}_{-0.2}$} & \edit1{$50.3^{+0.3}_{-0.3}$} & \edit1{$90.1^{+0.1}_{-0.1}$} & \edit1{$89.7^{+0.2}_{-0.2}$} & \edit1{$90.2^{+0.2}_{-0.2}$} \\
\enddata
\tablecomments{All fits share the identical data, spectral response, and sampler configuration; only the polarization response differs. The GEANT4-response fits are the production fits of the responses delivered with the data; the mode 1 and mode 2 fits are seeded, with the seeds recorded in the result files. The responses are built at $\delta=27.25^\circ$, $\lambda=6.07^\circ$. Differences quoted in the text are computed from the unrounded medians.}
\end{deluxetable*}

\section{Future Implementations and Conclusions}\label{sec:future}
\edit1{The results of this work are the} off-axis \edit1{azimuthal scattering angle distribution described by \cite{fabio} evaluated without approximation across the band of Section \ref{sec:intro} and extended by the effect of the detection threshold, two response modes built on it, and the software that builds them. The description carries the off-axis} projection effects that \edit1{most GRB polarization analyses to date have neglected. Mode 1 is built from the unpolarized simulations alone and needs about} 60 \edit1{times fewer simulations than} the GEANT4-based approach used for the POLAR catalog \citep{Kole2020}, with the small, characterized caveats discussed in Section \ref{sec:fakeGRBResults}. \edit1{Mode 2 adds fully polarized simulations at two polarization angles per energy, models the polarization-dependent instrumental effects from them, and matches the full GEANT4-response measurement from the faintest to the brightest cases tested (Section \ref{sec:mode2Results}). The released package, MENHAG (Modulation ENgine for Harmonic ASAD Generation), builds either response for any planar Compton polarimeter from that instrument's own simulations (Appendix \ref{app:menhag}).}

As the polarization response depends on the source incidence angle, the analytical framework makes it practical to generate a dense grid of responses across the full field of view of a detector. For POLAR this was computationally prohibitive as each per-GRB i.e., per incidence location, responses required approximately two weeks on a computing cluster. That is reduced to few hours when employing the analytical method presented here. The method was therefore already leveraged for preliminary analysis of the scientific prospects of the POLAR-2 mission \citep{Kole2025POLAR2}. This work also opens the door to time-resolved polarization analysis of long GRBs whose incidence angle changes significantly during the prompt emission. Dividing the burst into time bins each with its own response provides a more accurate probe of the polarization angle evolution. This provides more statistically significant constraints on the jet physics and emission mechanism of the prompt emission of GRBs \citep{Burgess2019, Kole2020}.  

\edit1{The} polarization-dependent systematics identified in Section \ref{sec:fakeGRBResults} \edit1{are addressed through the two-angle mode which models them from polarized simulations at two angles per energy, removing the polarization-degree bias of the unpolarized-only construction at all fluences probed. For future response-simulation campaigns we therefore recommend including fully polarized simulations at two polarization angles per energy, in addition to the unpolarized simulation, which enables the two-angle construction at any statistics level. Because mode 1 assumes that the instrument modifies polarized and unpolarized ASADs identically (Equation \ref{eq:assump}), the difference between the mode 2 and mode 1 responses of a mass model maps the part of the instrumental response that depends on the polarization state, a potential diagnostic for the design and calibration of future instruments.}

Beyond the POLAR mission, the framework developed here generalizes naturally to other wide field of view hard $X-$ray and $\gamma-$ray polarimeters, including AstroSAT-CZTI \citep{astrosat1} and the upcoming Daksha \edit1{\citep{Bhalerao2024Daksha}} mission. Compton imagers like COSI \citep{Tomsick2019COSI} and GRAPE \citep{OnateMelecio2024GRAPE} could also benefit from this approach by considering the polarization model in the frame of the photon without the transformation discussed in Section \ref{subsec:coordTrans}. Additionally, narrow field-of-view hard $X-$ray polarimeters such as XL-Calibur \citep{Abarr_2021} could benefit from this approach by constructing responses \edit1{at fine polarization-angle binning at no additional simulation cost}, as needed, for polarization measurements. 

\edit1{We} have presented a generalized description of the azimuthal Compton scattering angle distribution for off-axis polarized sources, applicable across \edit1{the hard X-ray and gamma-ray band of Compton polarimeters, from tens of keV to about 1 MeV, together with} rapid response-generation \edit1{methods that enable} practical polarization analysis of polarimeter data. The \edit1{two-angle mode} reproduces the results of full Monte-Carlo simulations \edit1{with two polarized simulations per energy instead of 60. The unpolarized-only mode requires none at all, reproduces them} for the overwhelming majority of GRBs observable by Compton polarimeter instruments\edit1{, and is the path of choice when polarized simulations are not attainable}.
 
\edit1{The response-generation methods described in this work are implemented in the released software package MENHAG, available at} \url{https://github.com/AstrophyJimmy/menhag} \edit1{and described in Appendix \ref{app:menhag}.}

\section*{Acknowledgments}

We thank the organizers and participants of the workshop \textit{"Advancing a Universal Analysis Framework for High-Energy Spectro-Polarimetry"} (Geneva, February 2025), supported by the Swiss National Science Foundation, for the fruitful discussions that helped refine aspects of this work. \edit1{M.G.}Z. acknowledges the support of the University of New Hampshire Department of Physics \& Astronomy through a teaching assistantship during the preparation of this work. 

\edit1{M.G.Z. acknowledges the use of Claude by Anthropic (Opus 5.5 and Fable 5 models) to assist with code packaging, code documentation improvements, and debugging tests for MENHAG. All of which were independently reviewed and scrutinized by the aforementioned author}.

\software{MENHAG (\citealt{zaid2026menhag}, version {1.0.0}), NumPy \citep{Harris2020NumPy}, SciPy \citep{Virtanen2020SciPy}, Matplotlib \citep{Hunter2007Matplotlib}, Astropy \citep{Astropy2013, Astropy2018, Astropy2022}, uproot \citep{uproot}, threeML \citep{3ML}, polpy \citep[][and its repository, \citealt{polarpy_repo}]{Burgess2019}, MultiNest \citep{Feroz2009MultiNest} through PyMultiNest \citep{Buchner2014PyMultiNest}}

\clearpage

\appendix
\section{Angle Definitions}\label{app:definitions}

The transformation from the photon frame to the detector frame introduces several angles. Table~\ref{tab:angles} collects their definitions for reference.
\startlongtable
\begin{deluxetable}{lll}
\tablecaption{Angles used throughout this work.\label{tab:angles}}
\tablehead{\colhead{Symbol} & \colhead{Name} & \colhead{Definition}}
\startdata
\cutinhead{Photon-frame angles ($x_\gamma y_\gamma z_\gamma$)}
$\theta\,(\theta_C)$ & Compton scattering angle    & Angle between $\vec{k}$ and $\vec{k}'$; polar angle from $-z_\gamma$ \\
$\theta_s$           & Spherical polar angle       & Standard polar angle from $+z_\gamma$; $\theta=\pi-\theta_s$ \\
$\phi$               & Azimuthal scattering angle  & Angle between $\vec{\xi}$ and the $xy$-projection of $\vec{k}'$, from $+x_\gamma$ \\
\cutinhead{Detector-frame angles ($xyz$)}
$\vartheta$          & Transformed polar angle     & Polar angle of $\vec{k}'$ in the detector frame; integration variable \\
$\varphi$            & Transformed azimuthal angle & Azimuthal angle of $\vec{k}'$ in the detector frame; integration variable \\
$\eta$               & Measured azimuthal angle    & Observable ASAD variable; $\eta=\phi+\phi_o$ for on-axis incidence \\
\cutinhead{Source incidence}
$\delta$             & Source zenith angle         & Incidence angle measured from the detector axis \\
$\lambda$            & Source azimuth angle        & Source azimuth in the detector frame \\
$\phi_o$             & Polarization angle          & (Local) angle of $\vec{\xi}$ from the detector $+x$-axis \\
\enddata
\tablecomments{A subscript $\gamma$ denotes a quantity in the incident-photon frame. Superscripts ``pol'' and ``unp'' denote the polarized and unpolarized beam, respectively.}
\end{deluxetable}
\section{\edit1{The Low-Energy Expansion of the Transformed Cross-Section}}\label{app:lowE}
\edit1{Following the approach of \cite{fabio}, the integrals in equation \ref{eq:integralsTransformed} are too complex to be evaluated analytically, especially in the transformed frame. However, the individual terms in the expanded KN cross-section (equation \ref{eq:expandedKN_Pol} and \ref{eq:expandedKN_Unp}) show a geometric series with the shape $(1-x)^{-n}$ where $x=-\varepsilon(1-\cos{\theta})$. So a first order approximation can be performed where $x\ll1$, which is the condition:}  
\begin{equation}\edit1{\label{eq:lowEnergyCond}
\varepsilon\ll1\rightarrow \frac{E}{511}\ll1\rightarrow E\ll511\;\mathrm{keV}}
\end{equation}
\edit1{At low energies of a few keV (typical X-ray energies), one can expand each term in equations \ref{eq:expandedKN_Pol} and \ref{eq:expandedKN_Unp} geometrically as $1+nx$, i.e.,} 
\begin{equation}\edit1{\label{eq:lowEnergyApprox}}
\begin{split}
    \edit1{\frac{1}{1+\varepsilon(1-\cos{\theta})}}&\edit1{\approx 1-\varepsilon(1-\cos{\theta})}\\
    \edit1{\frac{1}{[1+\varepsilon(1-\cos{\theta})]^3}}&\edit1{\approx 1-3\varepsilon(1-\cos{\theta})}\\
    \edit1{\frac{\sin^2{\theta}}{[1+\varepsilon(1-\cos{\theta})]^2}}&\edit1{\approx \sin^2{\theta}\big(1-2\varepsilon(1-\cos{\theta})\big)}\\
    \edit1{\frac{\sin^2{\theta}\cos^2{\phi}}{[1+\varepsilon(1-\cos{\theta})]^2}}&\edit1{\approx \sin^2{\theta}\cos^2{\phi}\big(1-2\varepsilon(1-\cos{\theta})\big)}\\
\end{split}
\edit1{}\end{equation}
\edit1{Then the transformation in equation \ref{eq:rotatingToPhotonFrame} is applied with the matrices as defined in equation \ref{eq:rotationMatrices} with $\lambda=\phi_o=0$ so this transformation would be limited to varying zenith angles $\delta$. The effect of varying the zenith angle $\delta$, changes the shape of the modulation curve from a single second harmonic, to multiple harmonics. Now, the transformation of the relevant terms becomes:} 
\begin{equation}\edit1{\label{lowEnergyApproxTerms}}
\begin{split}
    \edit1{\sin^2{\theta}\cos^2{\phi} =x^2_\gamma} &\edit1{= (\sin{\vartheta}\cos{\varphi}\cos{\delta}+\cos{\vartheta}\sin{\delta})^2}\\
    \edit1{\sin^2{\theta} =x^2_\gamma+y^2_\gamma} &\edit1{= 1-(\sin{\vartheta}\cos{\varphi}\sin{\delta}-\cos{\vartheta}\cos{\delta})^2}\\
    \edit1{\cos{\theta} =-z_\gamma} &\edit1{=-(\sin{\vartheta}\cos{\varphi}\sin{\delta}-\cos{\vartheta}\cos{\delta})}\\
\end{split}
\edit1{}\end{equation}
\edit1{Substituting the transformed terms from these equations into the approximated terms in equation \ref{eq:lowEnergyApprox} we can therefore evaluate the integrals in equation \ref{eq:integralsTransformed} analytically to find:} 
\begin{equation}\edit1{\label{eq:lowEnergyApproExpression}}
\begin{split}
\edit1{\mathcal{D}^{pol}(\varphi, \varepsilon, \delta)} &\edit1{= -\frac{9\varepsilon \cos^2 \delta \sin \delta}{32(2\varepsilon - 1)} \cos^3 \varphi - \frac{\cos^2 \delta}{2\pi} \cos^2 \varphi + \frac{9\varepsilon(\cos^2 \delta + 1) \sin \delta}{32(2\varepsilon - 1)} \cos \varphi + \frac{2 + \cos^2 \delta}{4\pi}}\\
\edit1{\mathcal{D}^{unp}(\varphi, \varepsilon, \delta)} &\edit1{= \frac{9\varepsilon \sin^3 \delta}{64(2\varepsilon - 1)} \cos^3 \varphi + \frac{\sin^2 \delta}{4\pi} \cos^2 \varphi + \frac{3\varepsilon(3 \cos^2 \delta + 4) \sin \delta}{64(2\varepsilon - 1)} \cos \varphi + \frac{\cos^2 \delta + 3}{8\pi}}
\end{split}    
\edit1{}\end{equation}
\edit1{Which provide expressions for the raw ASAD distribution without normalization for polarized and unpolarized sources as a function of the source energy $\varepsilon$ and zenith angle $\delta$ (the symbol $\mathcal{D}$ is reused here for the unnormalized azimuthal distribution; elsewhere it denotes the cross-section per solid angle). The normalization step in Equation \ref{eq:integrals} is essential only when constructing an ASAD for a spectrum. Each ASAD energy will correspond to a different normalization factor depending on the value of the integral in the denominator which can also be evaluated analytically under the same approximation. At face value, the distributions of Equation \ref{eq:lowEnergyApproExpression} are sufficient for studying mono-energetic sources. These equations clearly show the presence of higher harmonics in both polarized and unpolarized ASADs, since both expressions can be rewritten to resemble the mathematical form of a Fourier series (by expanding all powers of cosines into different harmonics, i.e., $A_n\cos(n\phi)$). This approach correctly predicted the shape of modulation curves for $X$-ray energies up to $\sim50\;\mathrm{keV}$  ($\varepsilon\approx0.1$) which was sufficient for X-ray polarimeters \cite{fabio}. However, at higher energies, the approximation in Equation \ref{eq:lowEnergyApproExpression} breaks down. As a result, energies exceeding $\sim 100\;\mathrm{keV}$ require a different approach.} 

\section{\edit1{The Threshold Gate as the Limit of Sampled Sharp Gates}}\label{app:gate}
\edit1{A real trigger turns on gradually, so we initially modeled it as a finite set of sharp thresholds. The thresholds are sampled at equal spacing from a Gaussian of mean $\mu$ and width $\sigma$; the gated ASAD of Equation \ref{eq:integralsTransformedStepFns} is evaluated at each sampled threshold, each curve is weighted by the Gaussian weight of its sample, and the weighted curves are summed. Samples at negative energies are assigned a sharp threshold of zero and keep their weight. This sampled construction remains available in the released code as a named option.}

\edit1{Letting the number of samples grow turns the weighted sum into an integral of the sharp-gated integrand over the threshold value $t$, with the Gaussian as the weight. A coincidence survives a threshold $t$ exactly when the smaller of the two deposits, $\min(E_e, E'_\gamma)$, exceeds $t$, so the sharp gate is a step function of that smaller deposit, and the integral of a step function against a Gaussian is the Gaussian cumulative distribution evaluated at the smaller deposit. The Heaviside gate inside the $\vartheta$ integrand is therefore replaced by the weight $w(\vartheta,\varphi)$ of Equation \ref{eq:erfGate}, which is the error-function gate adopted in Section \ref{sec:threshold}. The closed form is the limit of the weighted sum of the gated integrands taken before normalization; the normalization of Equation \ref{eq:integralsTransformedStepFns} is then applied once to the gated integrals.}

\section{\edit1{The MENHAG Software Package}} \label{app:menhag}

\edit1{The framework presented in this work is released as an open-source Python package, MENHAG (Modulation ENgine for Harmonic ASAD Generation), available at} \url{https://github.com/AstrophyJimmy/menhag} \edit1{and archived on Zenodo (DOI 10.5281/zenodo.23204599, version 1.0.0). The package computes the fully polarized azimuthal scattering angle distributions of a wide field-of-view planar Compton polarimeter from a small set of simulated inputs, in the two modes of this paper. Mode 1 takes, for each simulated energy, the unpolarized ASAD of the instrument together with the source incidence angles $(\delta,\lambda)$ and the threshold gate parameters $(\mu,\sigma)$ of Section \ref{sec:threshold}, and multiplies the unpolarized ASAD by the analytically computed modulation kernel of Equation \ref{eq:appSameIMinusNorm}, renormalized as in Equation \ref{eq:appSameIcomplete}, for every requested energy, polarization fraction and polarization angle; the gate values are instrument-specific inputs that the user measures or fits, and the sampled mixture of sharp gates of Appendix \ref{app:gate} remains available as an option. Mode 2 takes the unpolarized ASAD and the fully polarized ASADs at a user-chosen list of polarization angles, two at minimum (default $0^\circ$ and $45^\circ$), and reconstructs the ASAD at any polarization angle and fraction through Equation \ref{eq:sparseBasis}, combining more than two angles by least squares per bin; its output container is identical to that of mode 1, so the two modes are interchangeable in downstream fitting. The statistical uncertainties of the input histograms are propagated to every output bin as described in Section \ref{sec:errProp}.}

\edit1{The package reads and writes the POLAR-style ROOT histogram layout, with FITS and CSV adapters, a documented template for any other instrument format that is adapted by editing two reader and writer functions, and an export to the \texttt{.prsp} polarization-response format read by the \texttt{polpy} plugin for threeML. A configuration file lists the energies, the incidence angles, the gate, the polarization grid, the mode and its input angles, and the file paths; the repository carries the test suite, a minimal configuration, and three executed notebooks that reproduce some of the constructions of this paper. Users of the package should cite this paper as the primary reference for the method and the archived software release through its DOI \citep{zaid2026menhag}; the repository carries a citation file that encodes both.}

\bibliography{sample631}{}
\bibliographystyle{aasjournal}

\end{document}